\documentclass[a4paper,twocolumn]{revtex4}

\usepackage{amsmath,amssymb}

\newcommand{\ham}{\hat{\mathcal{H}}} 
\newcommand{\ii}{\mathrm{i}} 
\DeclareMathOperator{\e}{e} 

\usepackage{graphicx,color}

\begin{document}
\title[Title]{How to directly observe Landau levels in driven-dissipative strained honeycomb lattices}
\author{Grazia Salerno, Tomoki Ozawa, Hannah M. Price and Iacopo Carusotto}
\affiliation{INO-CNR BEC Center and Department of Physics, University of Trento - via Sommarive 14 38123 Povo, Italy}
\begin{abstract}
We study the driven-dissipative steady-state of a coherently-driven Bose field in a honeycomb lattice geometry. In the presence of a suitable spatial modulation of the hopping amplitudes, a valley-dependent artificial magnetic field appears and the low-energy eigenmodes have the form of relativistic Landau levels. We show how the main properties of the Landau levels can be extracted by observing the peaks in the absorption spectrum of the system and the corresponding spatial intensity distribution. Finally, quantitative predictions for realistic lattices based on photonic or microwave technologies are discussed.
\end{abstract}
\maketitle
\section{Introduction}
Artificially-prepared honeycomb lattices offer insight into parameter ranges that are out of reach for natural graphene, suggesting new directions in which to push experimental investigations in materials science~\cite{Polini}. One of the advantages of such artificial graphene is that the tunnelling of particles between different lattice sites can be controlled in an independent and flexible way.

A beautiful result from the theory of electron motion in honeycomb lattices is that the effect of an elastic deformation of the material, e.g. as induced by a mechanical strain, can be described in terms of an artificial valley-dependent magnetic field acting on the electrons.
This effect was first predicted in carbon nanotubes~\cite{KaneMele, Suzuura} and soon extended to two-dimensional graphene~\cite{Guinea1,deJuan,Guinea,Voz,Castroneto,Goerbig}. A suitably designed strain can generate a constant magnetic field such that the electronic spectrum exhibits quantized Landau levels, as experimentally observed in graphene nanobubbles~\cite{Levy}. As a consequence of the underlying Dirac dispersion of the electron band structure, the spacing of the Landau levels follows a square-root law starting from zero energy for both positive and negative energies, with the sign of the artificial magnetic field opposite for electrons in the vicinity of the two inequivalent Dirac points.

In the photonic context, this physics was first experimentally investigated in~\cite{Rechtsman} using a distorted lattice of optical waveguides: the lattice deformation modulated the hopping amplitude between neighbouring waveguides, which in turn generated an artificial pseudo-magnetic field for photons. Given the propagating nature of the optical set-up, evidence of the Landau levels could only be obtained in an indirect way from the localized edge modes, which reside in the energy-gaps between the Landau levels. In this experiment, it was not possible to obtain detailed information on the microscopic structure of the Landau levels themselves.

In this paper, we propose an alternative photonic set-up consisting of an array of cavities with a honeycomb geometry, as inspired by the experiments of~\cite{Amo1} and~\cite{Bellec1,Bellec2,Bellec3}: the former used a laterally patterned semiconductor microcavity device for infra-red light, while the latter were based on an array of coupled dielectric cylindrical resonators sandwiched between metallic plates. In contrast to the propagating waveguide set-up of~\cite{Rechtsman}, such cavity arrays are intrinsically driven-dissipative systems, which allow the use of spectroscopic techniques to characterize their eigenstates. In fact, a general result~\cite{RMP} is that the different eigenmodes of a driven-dissipative system naturally appear as peaks in the transmission/absorption spectra under a coherent incident field. When the pump frequency is set on resonance with a given mode spectrally isolated from its neighbours, the intensity profiles in both real- and momentum-space faithfully reproduce the mode wavefunction~\cite{RMP}.

Here we show how this general spectroscopic technique can be applied to the case of a honeycomb lattice to investigate the Landau levels appearing in the presence of distortion. Instead of a complex trigonal distortion of the lattice, as considered, for example, in~\cite{Rechtsman, Poli, Schomerus}, we consider a simpler geometry with a single hopping amplitude that is linearly varied~\cite{Gopalakrishnan}. Even though this configuration is hard to realise by mechanically straining a real graphene sample, for the sake of simplicity we shall refer to it as the {\em strained} honeycomb lattice.

The structure of the article is the following: in Sec.~\ref{sec:model} we review the tight-binding model of the honeycomb lattice, we introduce a configuration that can generate an artificial pseudo-magnetic field and we analytically study the resulting Landau levels. In Sec.~\ref{sec:exactD}, the analytically calculated Landau levels are compared with numerical diagonalization of the tight-binding model. The steady-state under a coherent driving is studied in Sec.~\ref{sec:drivendiss}: clear signatures of the Landau levels are highlighted and related to the analytically calculated mode energies and mode wavefunctions. In Sec.~\ref{sec:NNN} we extend our model to include next-nearest-neighbour (NNN) hoppings showing that the key features of pseudo-Landau levels remain unchanged. The experimental feasibility of our proposal is discussed in Sec.~\ref{sec:experim} using realistic parameters from available experiments. Finally, conclusions are drawn in Sec.~\ref{sec:conclu}.

\section{The model}
\label{sec:model}
\subsection{The honeycomb lattice}
\begin{figure}[b]
\centering
\includegraphics[width=0.4\textwidth]{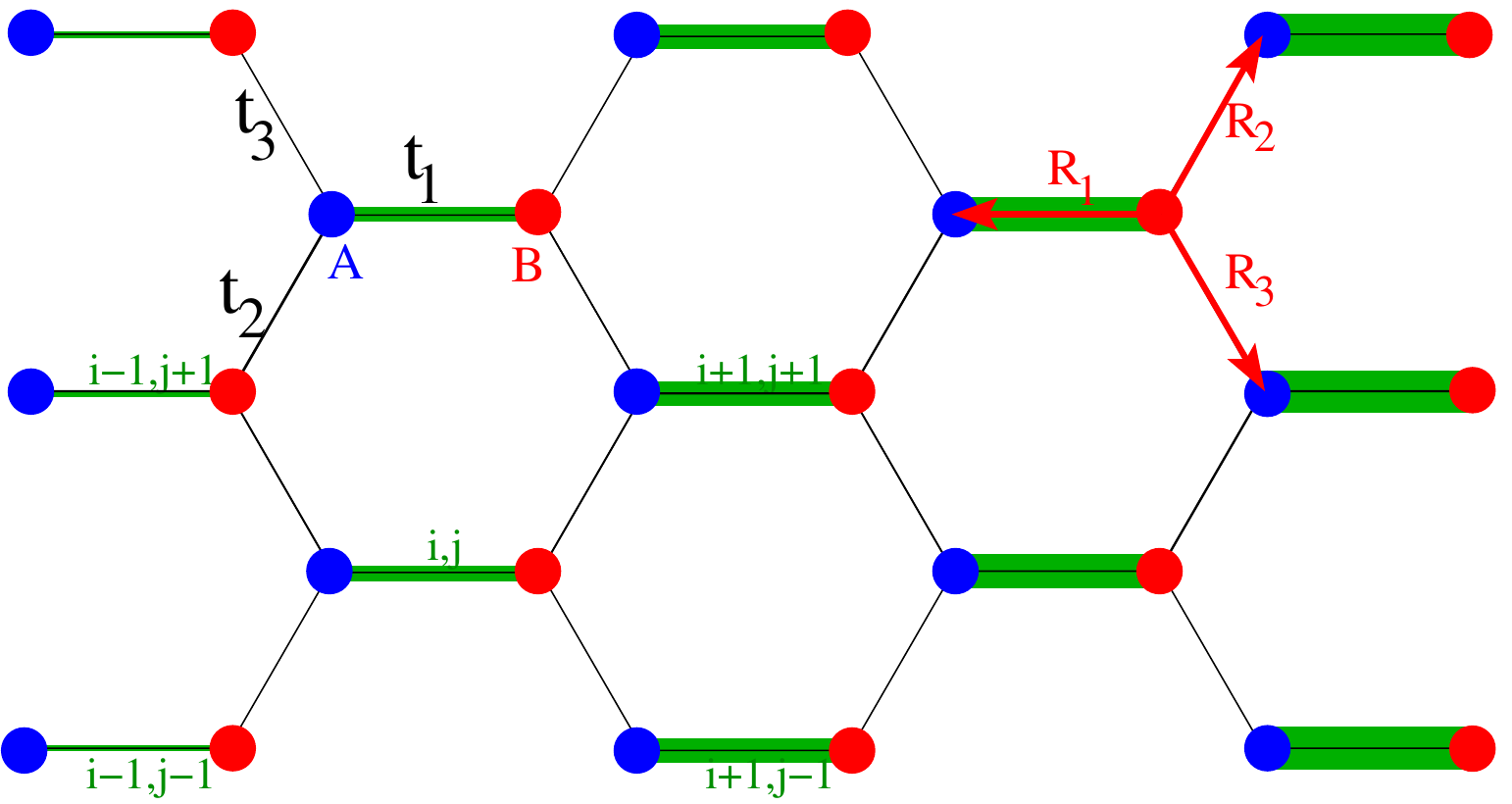}
\caption{The \textit{strained} honeycomb lattice, with $N_x$ unit cells along the $x$-armchair direction and $N_y$ along the $y$-bearded direction (in the figure $N_x=N_y=5$). A unit cell contains two inequivalent lattice points $A$ and $B$, separated by a unit lattice-spacing $a$. The unit cell is labelled with two indexes $(i,j)$. The nearest neighbour hopping strength along the vector $\vec{R}_l$ is denoted by $t_l$. The gradient of hopping $t_1^i$ along the armchair direction is schematically indicated by the thickness of the green line connecting $A$ and $B$ sites: the thicker the line, the stronger the coupling.}
\label{fig:graphene}
\end{figure}

In the honeycomb lattice, there are two inequivalent sites in each real-space unit cell. We label these sites as $A$ and $B$, and specify the nearest neighbour vectors as: $\vec{R}_1=\left(-a,0\right)$, $\vec{R}_2=\left(a/2,\sqrt{3}a/2\right)$ and $\vec{R}_3=\left(a/2,-\sqrt{3}a/2\right)$, where $a$ is the lattice spacing. Within a tight-binding approach, we first restrict our attention to nearest-neighbour hoppings of strength $t_l$ along each lattice vector $R_l$. Later on in Sec.~\ref{sec:NNN} this formalism will be extended to also include next-nearest-neighbour hoppings.

In a perfect honeycomb lattice, the nearest neighbour hoppings are spatially independent and equal $t_1=t_2=t_3$.
In the $A$ and $B$ sublattice basis, the wavefunction takes the form of a spinor, that can be interpreted as a \textit{pseudo-spin}. Setting the bare site energy as the zero energy, the bulk Hamiltonian in momentum space has the form:
\begin{equation}
\ham=\begin{pmatrix}
0     & V^*(\vec{k}) \\
V(\vec{k}) &  0
\end{pmatrix}
\label{tbh}
\end{equation}
with an off-diagonal matrix term: 
\begin{equation}
V(\vec{k})=t_1 \e^{\ii \vec{k}\cdot \vec{R}_1}+t_2 \e^{\ii \vec{k}\cdot \vec{R}_2}+t_3 \e^{\ii \vec{k}\cdot \vec{R}_3}.
\label{vk}
\end{equation} 
By diagonalizing the Hamiltonian in Eq.~\eqref{tbh}, we obtain the well-known band structure $\mathcal{E}(\vec{k})$ of the unstrained honeycomb lattice, which consists of two bands, symmetrically located with respect to the energy zero~\cite{Castroneto}.
 
The first Brillouin zone can be taken in the form of a hexagon and is delimited by the Dirac points $K$ and $K'$. The three equivalent $K$ points are located at $K=\left(\pm\frac{2\pi}{3a},\frac{2\pi}{3\sqrt{3}a}\right)$, $K=\left(0,-\frac{4\pi}{3\sqrt{3}a}\right)$ while the three equivalent $K'$ points are located at $K'=\left(\pm\frac{2\pi}{3a},-\frac{2\pi}{3\sqrt{3}a}\right)$, $K'=\left(0,\frac{4\pi}{3\sqrt{3}a}\right)$. At these six special points, the two bands touch at zero energy with a linear dispersion. 

\subsection{Strained honeycomb lattices}
\label{subsec:strain}
In real graphene, strain is mechanically generated by applying suitably designed external forces to the sample, so that atoms are physically pushed closer together or pulled further apart in a spatially-dependent way. This results in a corresponding modulation of both the sample geometry and the hopping amplitudes~\cite{Castroneto}. 
Typical geometries used in graphene experiments involve graphene nanobubbles~\cite{Levy} or trigonal deformations of the honeycomb lattice~\cite{Guinea}.
Such trigonal deformations were implemented in the photonic context in the coupled waveguide experiment of~\cite{Rechtsman}. In the present work, we take advantage of the wider design flexibility of artificial graphene to show how much simpler geometries with a unidirectional hopping gradient can be used to generate an artificial magnetic field.

To understand how a magnetic field naturally appears in a strained honeycomb lattice, one can follow the available literature~\cite{Castroneto} and make use of an extended tight-binding model with 	a suitably designed spatial dependence of the hopping amplitudes $t_l^{(i,j)}$ on the site indices $(i,j)$.

As a first step, we consider constant, spatially independent $t_1\neq t_2 \neq t_3$. 
We expand the Hamiltonian in Eq.~\eqref{tbh} around the $K,K'$ Dirac points by setting $\vec{k}=\left(q_x, q_y-\xi 4\pi/(3\sqrt{3}a)\right)$, where $|\vec{q}|\ll 1/a$ and the valley index $\xi=\pm 1$ labels the two inequivalent $K,K'$ Dirac points. At first order in $q$, we get:
\begin{equation}
\begin{split}
V(\vec{q})\approx & \,q_x \left(\frac{\sqrt{3}a}{4}\xi(t_2-t_3)-\frac{\ii a}{4}\left(4 t_1 +t_2 +t_3\right)\right)\\+& q_y \left(\frac{3a}{4}\xi (t_2+t_3)-\ii \frac{\sqrt{3}a}{4}(t_2-t_3)\right)\\ +&\frac{1}{2}(2t_1-t_2-t_3)+\ii\frac{\sqrt{3}}{2}\xi(t_2-t_3).
\end{split}
\label{vka}
\end{equation}

With a simple manipulation, we can re-write Eq.~\eqref{vka} in the form:
\begin{equation}
V(\vec{q})\approx -\ii v^x_D (\hbar q_x + e A_x) + v^y_D (\hbar q_y + e A_y)
\end{equation}
where the Dirac velocity is generally no longer isotropic,
\begin{equation} 
\begin{cases}
v^x_D = \frac{a}{4\hbar}\left(4 t_1 +t_2 +t_3\right)+\ii\frac{\sqrt{3}a}{4\hbar}\xi(t_2-t_3)
\\
v^y_D = \frac{3a}{4\hbar}(t_2+t_3)-\ii\frac{\sqrt{3}a}{4\hbar}\xi(t_2-t_3)
\end{cases}
\label{velocity}
\end{equation}
and an artificial magnetic vector potential appears of components $A_{x,y}$ such that:
\begin{equation} 
\begin{cases}
v_D^x e A_x = \frac{\sqrt{3}}{2}\xi \left(t_2-t_3\right)
\\
v_D^y e A_y= \frac{1}{2}\xi \left(2t_1-t_2-t_3\right).
\label{A}
\end{cases}
\end{equation}

In the $t_1=t_2=t_3$ case of an unperturbed honeycomb lattice, the artificial vector potential is obviously zero and the Dirac velocities $v_D^{x,y}$ in both two $x,y$ directions are equal to:
\begin{equation}
v_D^x=v_D^y=v_D=3at/2\hbar
\end{equation}

For generic, but still spatially uniform hoppings, $t_1\neq t_2 \neq t_3$, the band dispersion is distorted. Provided that the hopping imbalance is not too strong, the Dirac cones are still present and the valley index $\xi$ keeps its meaning, but the position of the Dirac cones in $\vec{k}$-space is shifted by the vector potential $\vec{A}$~\cite{Greif, Lim}. At the same time, the different values of the Dirac velocities in the $x,y$ directions $v_D^{x}\neq v_D^{y}$ and the appearance of an imaginary part in these signal that neither the group velocity nor the pseudo-spin are parallel to the wavevector $\vec{q}$ anymore. Moreover, this imaginary part can be understood as off-diagonal components in the Dirac velocity recast in a tensorial form~\cite{deJuan}.

In order to have a non-zero artificial magnetic field, one must allow for a spatial dependence of the hopping elements $t_l^{(i,j)}$ on the site indices $(i,j)$, as sketched in Fig.\ref{fig:graphene}: the position of the (local) Dirac points in $\vec{k}$ space then varies along the lattice in real space, following the spatial dependence of the artificial vector potential $\vec{A}$. The artificial magnetic field $B$ (directed along the $z$ direction) is then naturally defined as the curl of $\vec{A}$ and is generally associated with a small spatial dependence of the Dirac velocity~\cite{deJuan}.

As this strain-induced artificial gauge field does not break time reversal symmetry, the vector potential $\vec{A}$ and the magnetic field $\vec{B}$ have opposite signs in the two $\xi=\pm 1$ valleys: in this sense, it is not a true magnetic field, but rather a \textit{pseudo-magnetic field}. In the next section, we shall review how a spatially homogeneous pseudo-magnetic field rearranges the continuous conical dispersion around the Dirac points into a series of discrete Landau levels.

\subsection{Landau levels in a strained honeycomb lattice}
We consider a ribbon of artificial graphene, infinite along the $y$-direction and of finite size $L_x$ along the $x$-direction, with $N_x$ unit cells oriented as in Fig.~\ref{fig:graphene}. Along $x$, the ribbon is terminated on both ends with bearded edges. We assume that the hopping $t_1^{(i,j)}$ is positive and has the following spatial dependence:
\begin{equation}
t_1^{(i,j)} \equiv t_1(x_i)=t\,\left(1+\frac{x_i}{3a} \,\tau \right)
\label{strain}
\end{equation}
with a gradient oriented in the $x$ direction. The positive dimensionless parameter $\tau$ quantifies the intensity of the spatially inhomogeneous strain and the positions $x_i \in (-L_x/2,L_x/2)$. We assume that the $t_{2,3}$ hoppings in the other two directions are spatially uniform and equal $t_2 = t_3 = t$. 

We choose the hopping in the middle of the ribbon to be $t_1(0)= t$. 
As we require the hoppings to have the same sign at all positions along $x$, the condition $t_1({-L_x/2})\geq0$ at one edge imposes an upper bound on the strain,
\begin{equation}
\frac{\tau L_x}{6a} < 1.
\label{cond0}
\end{equation}
This condition automatically implies that $t_1(L_x/2)\leq 2 t$ at the other edge, which guarantees (within a local band picture) that no local Lifshitz transition to a gapped state~\cite{Montambaux,Bellec2} takes place in the considered ribbon.

The form of the strain in Eq.~\eqref{strain} leads to a vector potential $\vec{A}$ in the Landau gauge with a constant valley-dependent pseudo-magnetic field $B=\partial_x A_y-\partial_y A_x$ of strength:
\begin{equation}
e B = \xi \frac{2 \hbar } {9 a^2} \tau.
\label{magneticfield}
\end{equation}
The corresponding magnetic length is:
\begin{equation}
l_B\equiv \sqrt{\hbar/|eB|}=3a/ \sqrt{2\tau}.
\end{equation}
Inserting the values of the electron charge and the lattice spacing of real graphene, the field in Eq.~\eqref{magneticfield} would correspond to a magnetic field $B\approx 2\tau 10^{3} $~T.

Inserting the specific form of strain Eq.~\eqref{strain} into Eq.~\eqref{velocity}, the two components of the Dirac velocity become:
\begin{equation}
\begin{cases}
v^x_D=v_D+\tau t \hat{x}/3\hbar \\
v^y_D=v_D.
\label{velocity2}
\end{cases}
\end{equation}
If the modulation of the hopping along the $x$ direction is small compared to the hoppings in the other two directions, we can begin our discussion by neglecting the spatial variation of the Dirac velocity. In more quantitative terms, the variation of $v_D^x$ on a magnetic length $l_B$ remains small compared to $v_D$ as long as the strain is weak $\sqrt{\tau}\ll 1$. Note that in the strained geometry considered in this paper the spatial dependence \eqref{velocity2} of the Dirac velocity has no impact on the vector potential amplitude $A_{x,y}$ given in Eq.~\eqref{A}.

The procedure to diagonalize the Hamiltonian is closely analogous to the usual one for charged massive particles in a uniform magnetic field in free space~\cite{Castroneto}. At the lowest order in $q$ and $x$, we can express the spatial dependence of the hopping amplitudes in terms of a position operator $\hat{x}$ and the off-diagonal matrix element in Eq.~\eqref{vka} becomes an operator,
\begin{equation}
\hat{V}=-\ii \hbar \hat{q}_x v_D +\xi\hbar \hat{q}_y v_D + t \tau \hat{x}/3a
\label{ho1}
\end{equation}
acting on the spatial wavefunction.
Using the standard canonical relation $[\hat{x},\hat{q}_x]=\ii $, it is straightforward to show that the commutator $[\hat{V}^\dagger,\hat{V}]= t^2 \tau$.
We can therefore introduce a creation operator $\hat{a}^\dagger \equiv \hat{V}/(t \sqrt{\tau})$ which satisfies the commutation rules $[\hat{a}, \hat{a}^\dagger] = 1$ of a quantum harmonic oscillator and recast Eq.~\eqref{ho1} into the form of the creation operator of a shifted harmonic oscillator:
\begin{equation}
\hat{V}=\hbar \omega \left(\sqrt{\frac{m \omega}{2 \hbar}}(\hat{x}-x_0) -\ii \sqrt{\frac{1}{2m\hbar\omega}} \hbar\hat{q}_x \right)=\hbar \omega \hat{a}^\dagger
\label{ho2}
\end{equation}
where $m \equiv t\sqrt{\tau}/(2v_D^2)$, the oscillation center is shifted in the $x$ direction by the $q_y$-dependent distance
\begin{equation}
x_0\equiv-\frac{3 \xi\hbar v_D a}{t\tau}\hat{q}_y=-\xi \,l_B^2 \,\hat{q}_y
\label{shift}
\end{equation}
and the oscillator frequency
\begin{equation}
\hbar \omega \equiv \sqrt{\frac{2\hbar v_D t \tau}{3a}}=t\sqrt{\tau}
\end{equation}
recovers the cyclotron frequency of a massless Dirac particle in a magnetic field,
\begin{equation}
\omega = v_D \sqrt{2 |e B|/\hbar}.
\end{equation}

To obtain the eigenfunctions $\phi_{A,B}(x,y)$, we now re-write the Hamiltonian in Eq.~\eqref{tbh} around the Dirac points as:
\begin{equation}
\ham 
\begin{pmatrix}
\phi_A \\
\phi_B
\end{pmatrix}
=\hbar \omega\begin{pmatrix}
0  & \hat{a}\\
\hat{a}^\dagger &0
\end{pmatrix}
\begin{pmatrix}
\phi_A \\
\phi_B
\end{pmatrix}=
\mathcal{E}
\begin{pmatrix}
\phi_A \\
\phi_B
\end{pmatrix}.
\label{eq0}
\end{equation}

Separating the equations, we have:
\begin{eqnarray}
(\hbar \omega)^2\hat{a} \hat{a}^\dagger \phi_A=&\mathcal{E}^2 \phi_A\\
\label{eq1}
(\hbar \omega)^2\hat{a}^\dagger \hat{a} \phi_B=&\mathcal{E}^2 \phi_B.
\label{eq2}
\end{eqnarray}
From Eq.(\ref{eq2}), we immediately see that $\phi_B$ is an eigenvector of the number operator: $\hat{a}^\dagger \hat{a} |N\rangle = N |N\rangle$ with an non-negative integer eigenvalue $N\geq 0$. Therefore, $\phi_B$ is a 1D harmonic oscillator eigenfunction $|N\rangle$ with frequency $\omega$, centred at the $q_y$-dependent position \eqref{shift}. Most remarkably, for each $\phi_B$ with a given $N\neq 0$, two independent eigenstates exist with opposite energies $\mathcal{E} = \pm\hbar \omega \sqrt{N}$.

Obtaining the corresponding $\phi_A$ requires a bit more care: from Eq.~\eqref{eq0}, for $N=0$ one finds a single eigenstate with $\phi_A=0$, while
\begin{equation}
\phi_A=\frac{\hbar \omega }{\mathcal{E}} \hat{a} |N\rangle=\frac{\hbar \omega}{\pm \hbar \omega \sqrt{N}}\sqrt{N} |N-1\rangle=\pm|N-1\rangle
\end{equation}
for $N>0$. 

To summarize, both the positive and negative energy states can be organized in a single sequence labelled by an integer $-\infty < n < \infty$ and their energies follow the square-root law of relativistic Landau levels with a cyclotron frequency $\omega$,
\begin{equation}
\mathcal{E}_n = \pm \hbar \omega \sqrt{|n|} =\pm t \sqrt{\tau\,|n|}.
\label{ll}
\end{equation}
For each eigenstate, the total wave function in the position representation is:
\begin{equation}
\psi_{\pm|n|}(x,y)=
\e^{\ii q_y y}\begin{pmatrix}
\phi_A(x) \\
\phi_B(x)
\end{pmatrix}=\e^{\ii q_y y}
\begin{pmatrix}
\pm \langle{x}||n|-1\rangle \\
\langle{x}||n|\rangle
\end{pmatrix},
\label{statell}
\end{equation}
where $|N\rangle$ can be explicitly written in the position representation as a Hermite polynomial of degree $N\geq0$, 
\begin{equation}
\langle x |N\rangle\propto \e^{-\left(x-x_0\right)^2/(2 l_B^2)} \mathrm{H}_N\left(\frac{x-x_0}{l_B}\right)
\label{statexll}
\end{equation}
and we have implicitly assumed that $|-1\rangle=0$. This full wavefunction $\psi(x,y)$ is therefore a spinor with a Landau level wave function in each component: for $|n|>0$, the relative sign of the two components $\phi_A$ and $\phi_B$ is opposite for opposite eigenstates at $\pm n$. For $n=0$ the state is completely localized on the $B$ sublattice.

Notice that there is no dependence on the valley index $\xi$ in the spinor of Eq.~\eqref{statell}; therefore the wave function in the strained lattice is the same for the two Dirac valleys. This is an important difference to the case of a non-strained system in a (real) external magnetic field, where the wave function is valley-dependent and the role of the $A$ and $B$ sublattices is inverted when passing from $\xi=1$ to $\xi=-1$ \cite{Goerbig}.

All the discussion so far neglects the dependence of the Dirac velocity on the hopping amplitudes in Eq.~\eqref{velocity}.
The first correction to the relativistic Landau levels in Eq.~\eqref{ll} comes from the spatial dependence of the Dirac velocity in Eq.~\eqref{velocity2}. Substituting the value $v_D^x\simeq v_D+ t\tau x_0/3\hbar$ evaluated at the oscillation center $x_0$ into \eqref{ho1}, one obtains an expression of the Landau levels that includes the first order correction,
\begin{equation}
\mathcal{E}_n=\pm t\sqrt{\tau\,|n|}\,\sqrt{1- \xi q_y a}.
\label{llcorrection}
\end{equation}
This correction shifts each level around the $K,K'$ points with a square-root dependence on $q_y$: the fact that the resulting levels are no longer flat is a key difference with respect to the standard Landau levels in the presence of a real magnetic field, where the Dirac velocity remains independent of position. Further corrections coming from second and higher order terms in the expansion \eqref{vka} of $V(\vec{q})$ in powers of $q$ go beyond the scope of the present work.

\section{Comparison with numerical results from exact diagonalization}
\label{sec:exactD}
\begin{figure}[]
\hspace*{-1em}
\includegraphics[width=0.51\textwidth]{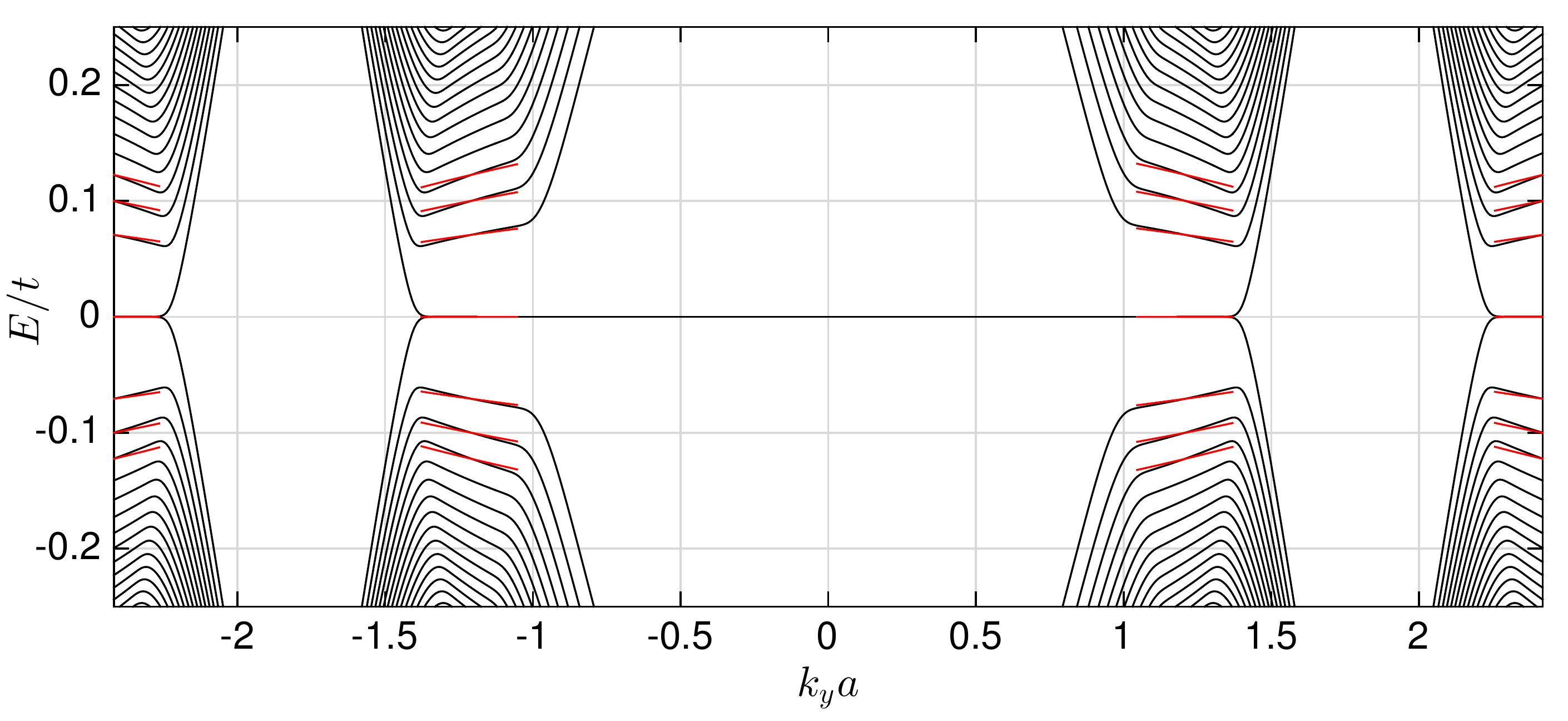}
\caption{The structure of levels around zero energy as a function of $k_y$, in units of the bare hopping $t$. This is numerically calculated from exact diagonalization of the tight-binding Hamiltonian for a ribbon of $N_x=601$ unit cells along $x$ with a strain of $\tau=0.005$, and periodic boundary conditions along the $y$-direction. Relativistic Landau levels  appear around the Dirac points $K$ and $K'$ for $k_y a\simeq \pm 1.21$ and $\pm 2.42$. Red lines indicates the analytical prediction for the lowest Landau levels including the first order correction to the Dirac velocity according to Eq.~\eqref{llcorrection}.}
\label{fig:ll}
\end{figure} 

The analytical results reviewed in the previous section were based on several approximations, in particular the very notion of an artificial vector potential $\vec{A}$ relied on a local band structure for each value of the hopping $t_l^{(i,j)}$. In order to verify the quantitative accuracy of this approach, we have numerically diagonalized the full tight-binding Hamiltonian and compared the outcome with the analytical approximation. 

An example of such a comparison is shown in Fig.~\ref{fig:ll} for the energy spectra of a large system of $N_x=601$ unit cells along $x$ with a relatively small strain parameter of $\tau=0.005$. As we are considering periodic boundary conditions along $y$, we can plot the spectrum as a function of the corresponding $k_y$.
The energy levels at zero energy are doubly degenerate, consisting of the $n=0$ Landau level and of the localized edge state associated with the bearded edges \cite{Kohmoto}. Around the Dirac points, we highlight the formation of quantized Landau levels. Their energies are in good agreement with the analytical prediction of Eq.~\eqref{llcorrection}, including the corrections due to the spatial dependence of the Dirac velocity $v_D^x$. The slight discrepancies that are visible to a careful eye can be explained by the approximations in our analytical calculations, e.g. the neglect of higher-order terms in the expansion \eqref{vka}.

As expected, the agreement between the analytical model and the full numerics gets better for smaller values of the strain parameter $\tau$: in this regime, the magnetic length extends for a larger number of sites (proportional to $1/\sqrt{\tau}$) and the $\vec{k}$-space wavefunction gets more localized in the vicinity of the Dirac point. This makes the continuum approximation underlying the analytical model more and more accurate. At the same time, the characteristic value of the vector potential $\vec{A}$ within the real-space wavefunction decreases as $\sqrt{\tau}$, thus reducing the importance of the corrections to the isotropic conical Dirac dispersion. Of course, this accuracy comes at the price of a reduced energy spacing of the Landau levels, again proportional to $\sqrt{\tau}$.

\begin{figure}[b]
\hspace*{-1em}
\includegraphics[width=0.5\textwidth]{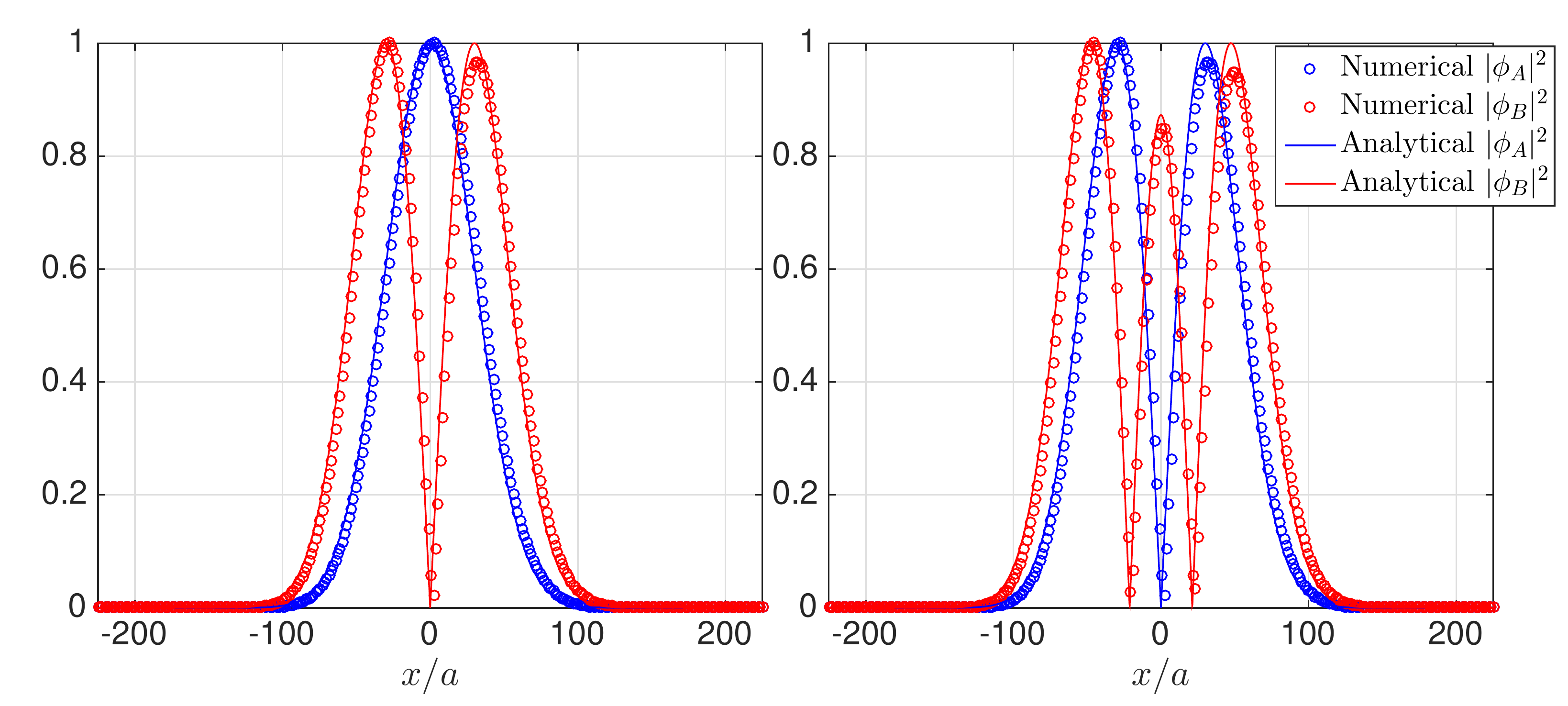}
\caption{Square modulus of the wavefunction of Landau levels around the $K'$ point as numerically obtained from exact diagonalization of the tight-binding Hamiltonian in Eq.~\eqref{tbh}, choosing $k_y=4\pi/(3\sqrt{3}a)$. The wave function has been separated for the two sublattices $|\phi_A|^2$ (blue) and $|\phi_B|^2$ (red). Left panel: $n=1$, right panel: $n=2$. Dots correspond to the numerical eigenfunction, lines to the analytical wavefunction in Eq.~\eqref{statexll}.}
\label{fig:n12ll}
\end{figure}

The spatial structure of the eigenfunctions is studied in Fig.~\ref{fig:n12ll}, where the square modulus of the numerical eigenfunctions is plotted for two Landau levels with $n=1$ and $n=2$, and the contributions from the $A$ and $B$ sites are separated. The left panel shows the states for $n=1$: according to Eq.~\eqref{statell}, the total wave function is $\psi_{n=1}\propto (|0\rangle,|1\rangle)$. On $A$ sites, this corresponds to a zero-th order Hermite polynomial, so the wave function has a Gaussian profile, as visible from the blue curve. On $B$ sites, instead, the wave function is proportional to a first-order Hermite polynomial, and has one node as visible from the red curve. In the right panel of Fig.~\ref{fig:n12ll}, which shows the $n=2$ Landau level, there is one node in the $A$-site wave function as expected, and two nodes in the $B$-site state. The slight asymmetry around the centre depends on the sign of the hopping gradient.

As predicted by the analytical model, exactly at the $K,K'$ points the wave function is symmetrically centred in the ribbon while, as $k_y$ moves away from the Dirac point, its center $x_0$ is shifted according to Eq.~\eqref{shift}. 
The finite extension in $k_y$ of the Landau level structure around the Dirac points, roughly indicated in Fig.~\ref{fig:ll} by the extension of the red lines, is limited by the Landau level touching the physical boundary of the system at $x=\pm L_x/2$. In the spectra, this is apparent as a sudden increase of the level energy.

\section{Steady state in a coherently driven dissipative lattice}
\label{sec:drivendiss}
After reviewing the main properties of Landau levels in a strained honeycomb lattice, we can now describe our proposal to probe the Landau spectra by studying the steady state of a driven-dissipative system, such as a photonic lattice made of a coupled cavity array\cite{Amo1} or microwave resonators \cite{Bellec1,Bellec2,Bellec3}. In both cases, the nearest neighbour hopping is due to the spatial overlap between modes localized on adjacent sites, so the required spatial dependence of the hopping Eq.~\eqref{strain} can be tailored by a careful design of the distance between neighbouring sites. Note that obtaining a gradient of $t_1$ along the $x$ direction does not involve distorting the lattice, but only varying the distance along neighbouring sites along $x$.

Damping naturally appears due to the unavoidable photon absorption and radiative losses, so a coherent driving by an external coherent light source is considered. A related approach was used in~\cite{Ozawa} to study anomalous and integer quantum Hall effects in a square lattice in the presence of a synthetic magnetic field and the anomalous quantum Hall effect in a honeycomb lattice with sub-lattice asymmetry.

We consider a large but finite honeycomb lattice, as sketched in Fig.~\ref{fig:graphene}. We assume that photons are lost uniformly from all sites at the rate $\gamma$ and the coherent pump is spatially localized in the central part of the lattice. This is beneficial to focus on the bulk properties of the lattice and to suppress spurious effects due to reflections from the lattice edges.

We coherently pump the system with a monochromatic field at frequency $\omega_0$. The pump has a spatial amplitude $f_{i,j}$ so that, in the steady state, the time dependent fields over $A$ or $B$ sites at position $(i,j)$ are:
\begin{equation}
a_{i,j}(T)=a_{i,j} \e^{-\ii\omega_0 T} \qquad b_{i,j}(T)=b_{i,j} \e^{-\ii\omega_0 T}.
\end{equation}
with the time-independent amplitudes $a_{i,j}$ and $b_{i,j}$ satisfying a linear system of Heisenberg equations~\cite{Ozawa}:
\begin{equation}
\begin{split}
&\hbar\left(\omega_0 +\ii \gamma\right) a_{i,j} + t_1^i \,b_{i,j}+ t \,b_{i-1,j-1} + t\, b_{i-1,j+1} = f_{i,j}\\
&\hbar\left(\omega_0 +\ii \gamma \right) b_{i,j} + t_1^i\, a_{i,j}+ t \,a_{i+1,j+1} + t\, a_{i+1,j-1} = f_{i,j}
\end{split}
\label{steadystate}
\end{equation}
for the indexing given in Fig.~\ref{fig:graphene}. The strong dependence of the non-equilibrium steady-state on the pump frequency $\omega_0$, as visible in \eqref{steadystate}, is the key feature allowing for the spectroscopic study of the eigenmodes. Conversely, waveguide experiments, such as the ones in~\cite{Rechtsman}, which are based on a conservative propagation equation, can hardly resolve the different eigenmodes.

As we want to probe Landau levels arising from strain, it is important to separate the contribution of the two inequivalent points $K$ and $K'$ as around these points the pseudo-magnetic field has opposite signs. To this end, we adopt the technique proposed in~\cite{Ozawa} and we assume the coherent driving to have a Gaussian spatial profile of width $\sigma_{x,y}$ in the two $x,y$ directions, and to have an in-plane wave vector in the vicinity of, e.g., the $K'$ Dirac point:
\begin{equation}
f_{i,j}= f_0 \exp\left(-\frac{x_{i,j}^2}{2 \sigma_x^2}\right) \exp\left(-\frac{y_{i,j}^2}{2 \sigma_y^2}\right) \e^{\ii \vec{K'}\cdot \vec{R}_{i,j}}
\label{pump}
\end{equation}
where $\vec{R}_{i,j}$ is the position vector of the appropriate site of the hexagonal lattice and the origin is assumed to be located in the middle of the central unit cell. Provided the spatial extensions $\sigma_{x,y} \gg a$, the coherent pump selectively addresses a small region in $\vec{k}$-space in the vicinity of the desired $K'$ Dirac point and efficiently excludes the other Dirac point.
\begin{figure}[]
\hspace*{-1em}
\includegraphics[width=0.24\textwidth]{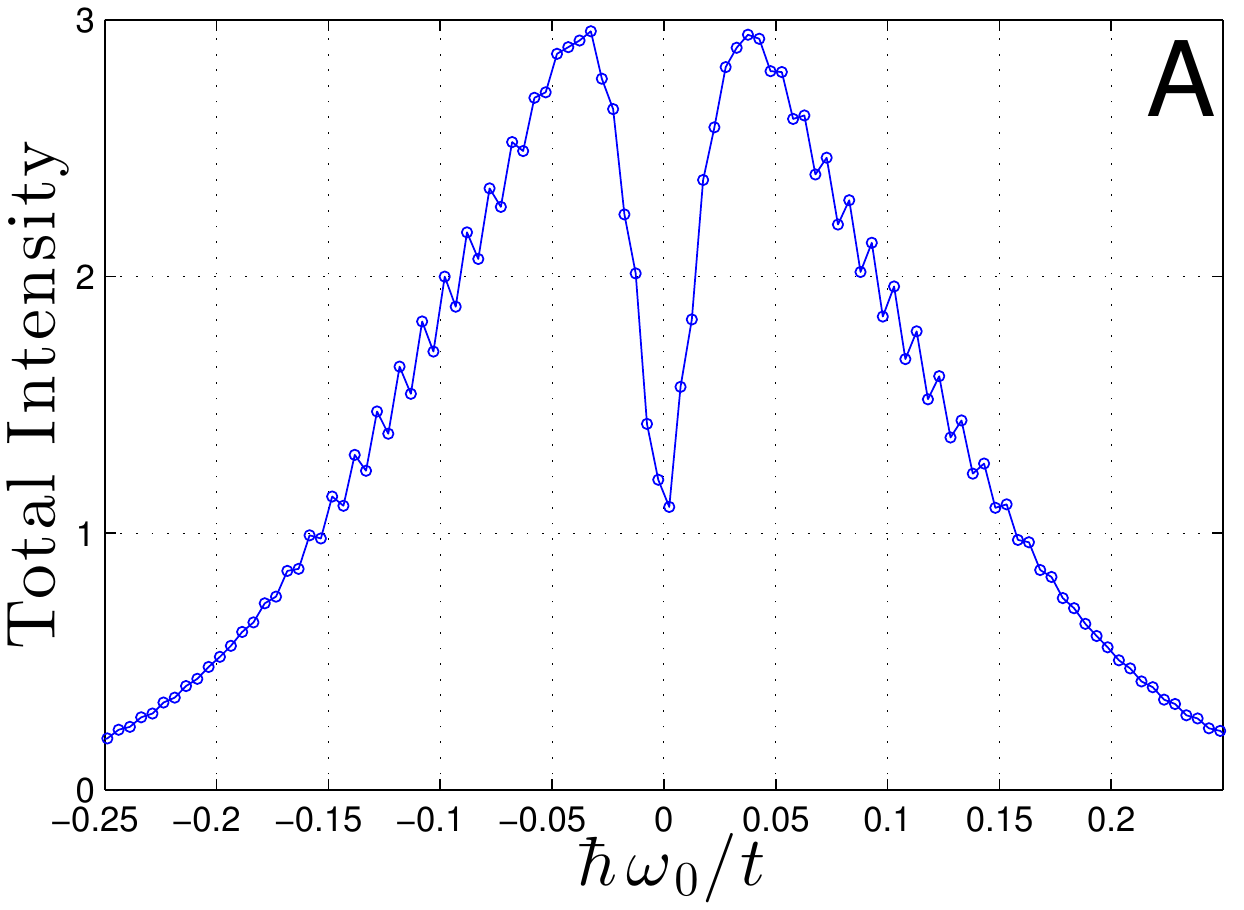}
\includegraphics[width=0.24\textwidth]{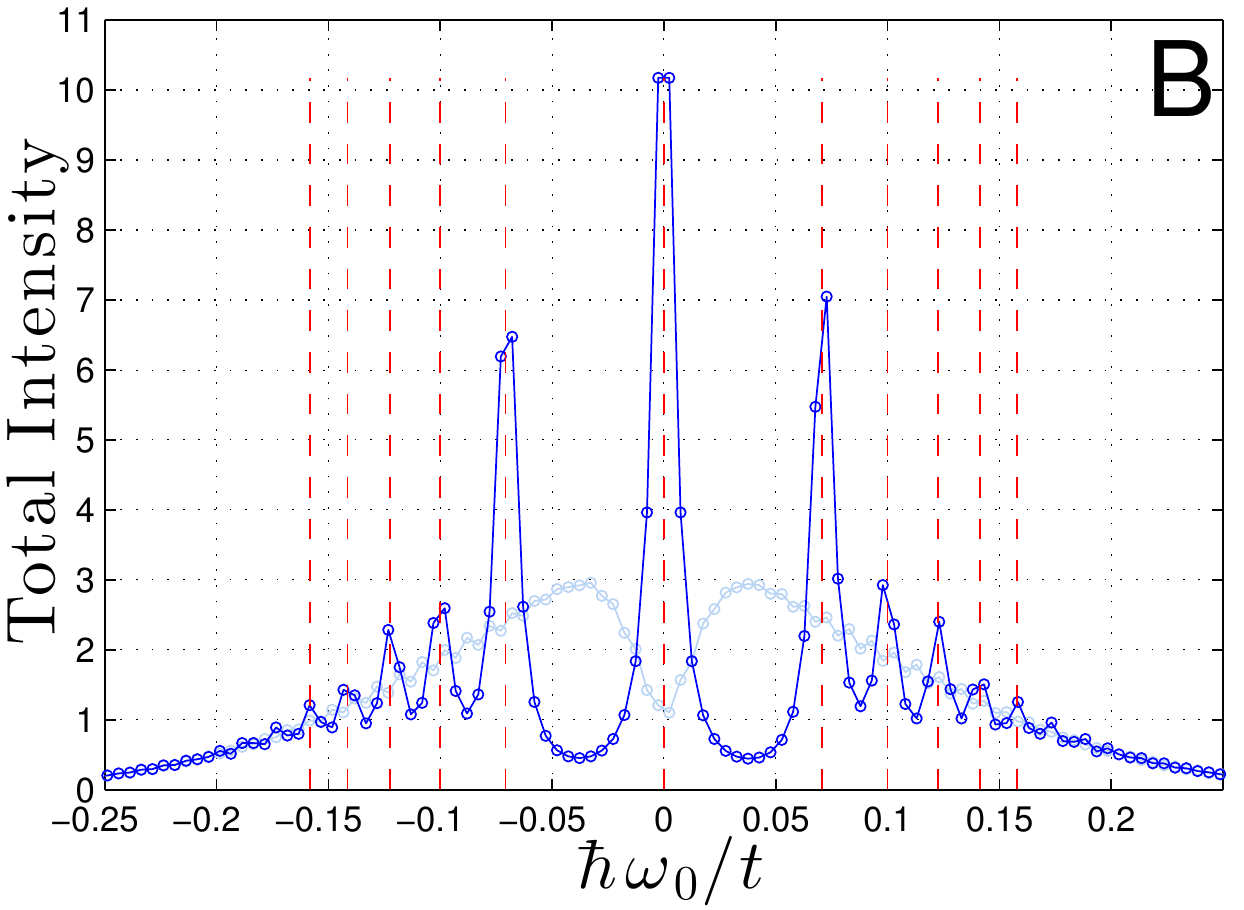}
\caption{Dots show numerically calculated spectra of the total field intensity in Eq.~\eqref{It} as a function of the pumping frequency $\omega_0$ for two values of the strain parameter, in arbitrary units. The loss rate is $\hbar\gamma/t=0.005$, and we take $N=601$ and $\sigma_x/a=\sigma_y/a=10$ for all panels. Panel \textbf{A} is for $\tau=0$ (no strain), while panel \textbf{B} is for $\tau=0.005$ (strained) in dark blue and $\tau=0$ in light blue. The solid lines are a guide to the eye and the red dashed lines indicate the analytical energies of the first six Landau levels as predicted by Eq.~\eqref{ll}.}
\label{fig:spectra}
\end{figure}

In Fig.~\ref{fig:spectra} we show examples of spectra of the total intensity summed over all lattice sites 
\begin{equation}
\label{It}
I_T=\sum_{i,j}\left(|a_{i,j}|^2+|b_{i,j}|^2\right)
\end{equation}
as a function of the pump frequency $\omega_0$. Depending on the specific configuration~\cite{RMP,Amo1,Bellec1,Bellec2,Bellec3}, such total intensity spectra can be experimentally accessed with either an absorption or a transmission measurement.

Panel~\ref{fig:spectra}\textbf{A} shows the spectra of the unstrained case $\tau=0$. In this case, the eigenstates of the honeycomb lattice~\cite{Castroneto} form a continuum with a conical Dirac dispersion $\omega\simeq v_D\,|q|$ in the vicinity of the $K,K'$ points, so the spectrum is a featureless continuum.  The small oscillations in the spectra are due to finite size effects stemming from a small but non-zero reflection at the edges and these disappear if larger systems are considered. 

When a small strain $\tau=0.005$ is introduced,  pronounced peaks emerge in the spectra shown in panel~\ref{fig:spectra}\textbf{B}, corresponding to the Landau levels. We compare the numerical spectra with the analytical energies of Landau levels, as given by Eq.~\eqref{ll}. We see that the position of the peaks in Fig.~\ref{fig:spectra} is in good agreement with the analytical values.

As usual, for a pump frequency close to resonance with a peak, the field intensity profile follows the wave function of the corresponding mode: the two cases of unstrained and strained honeycomb lattices will be separately discussed in the next sections.

\begin{figure}[h!]
\hspace*{-0.5em}
\includegraphics[width=0.24\textwidth]{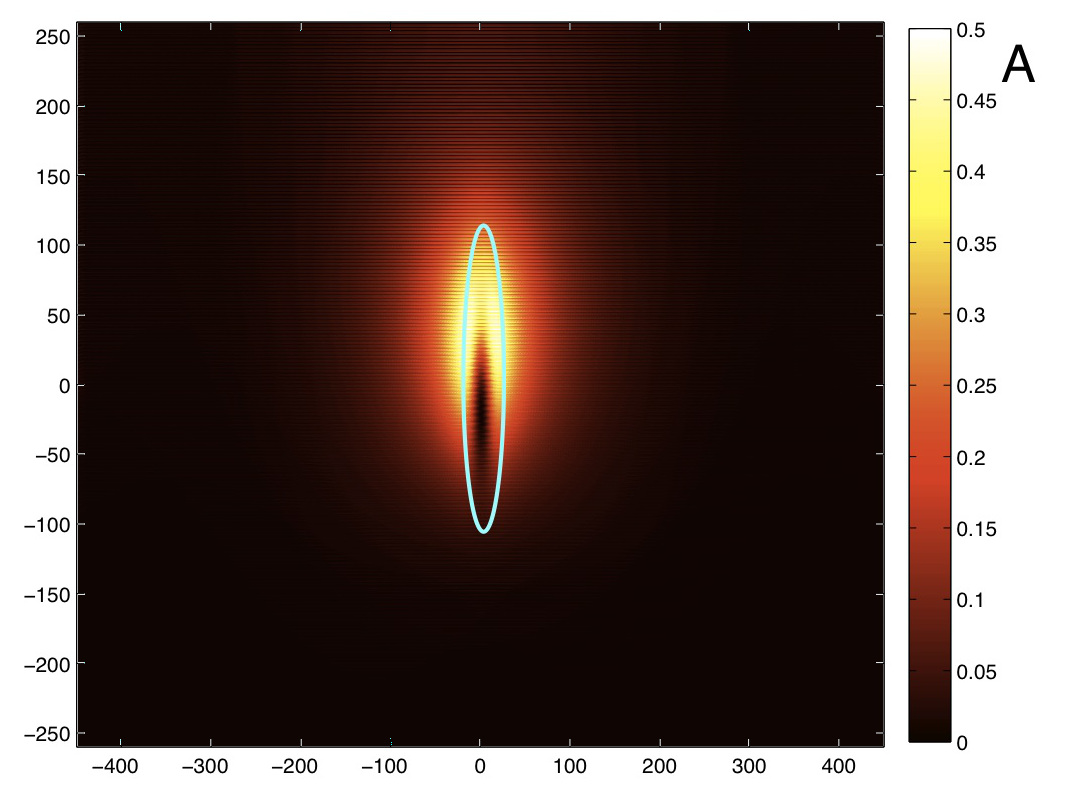}
\includegraphics[width=0.24\textwidth]{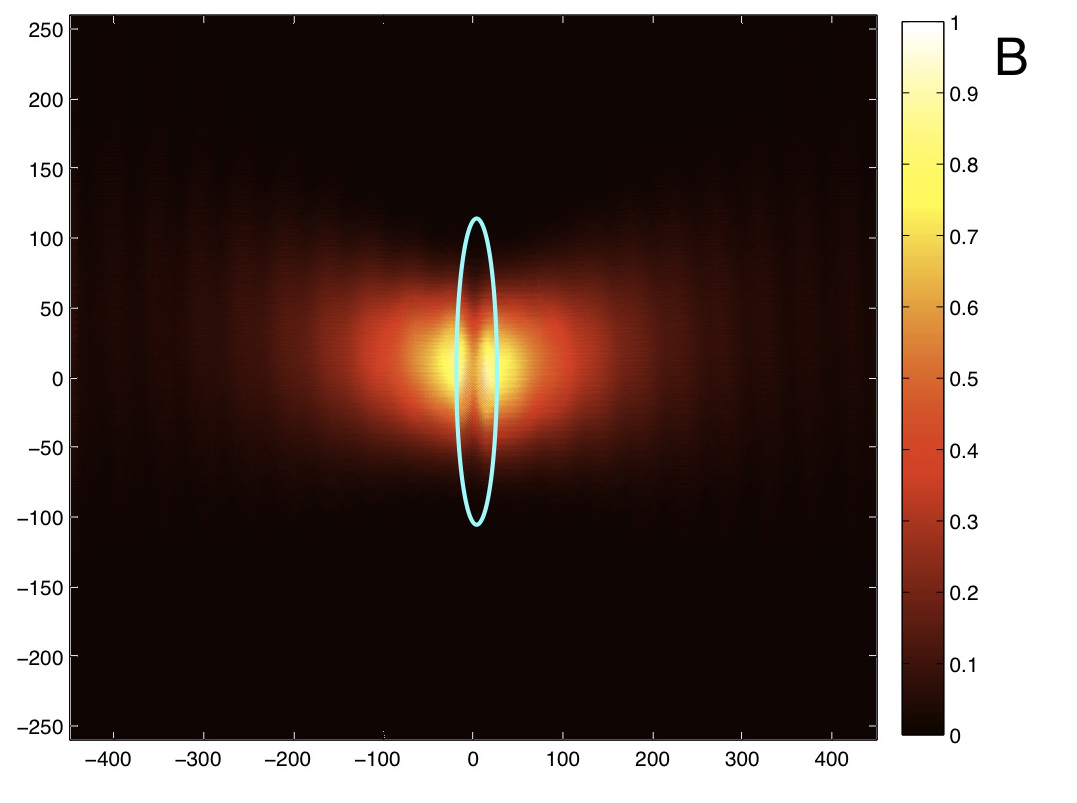}
\hspace*{-0.5em}
\includegraphics[width=0.24\textwidth]{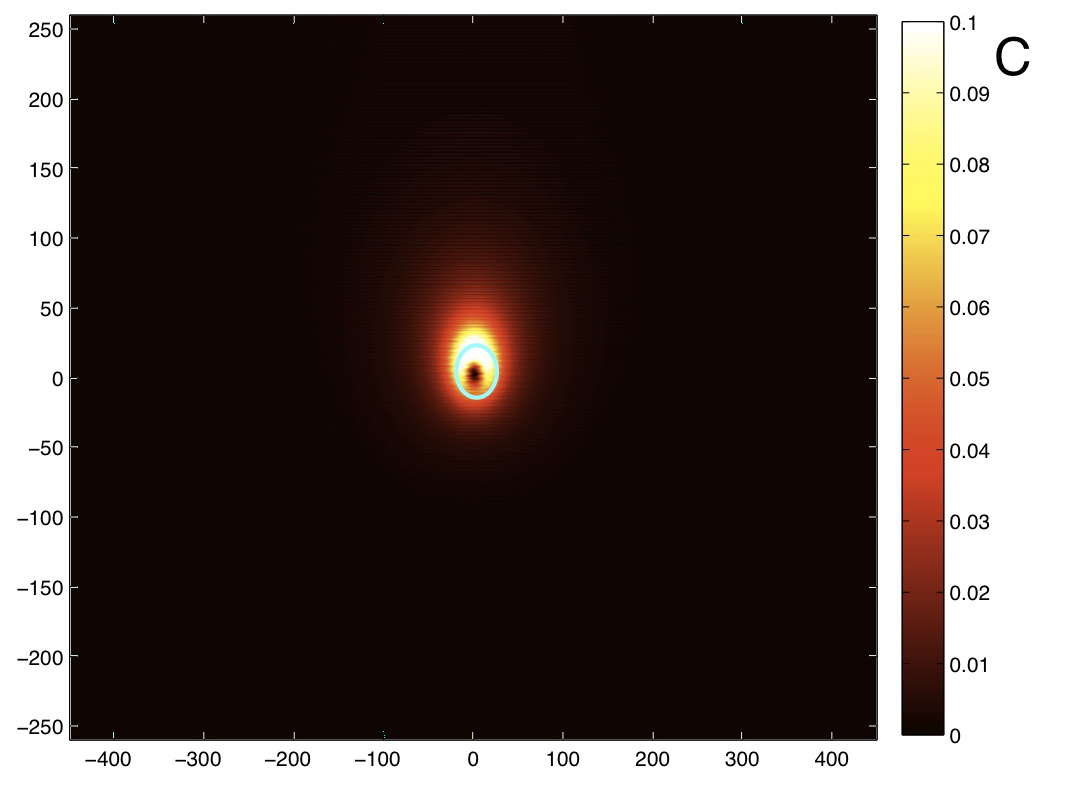}
\includegraphics[width=0.24\textwidth]{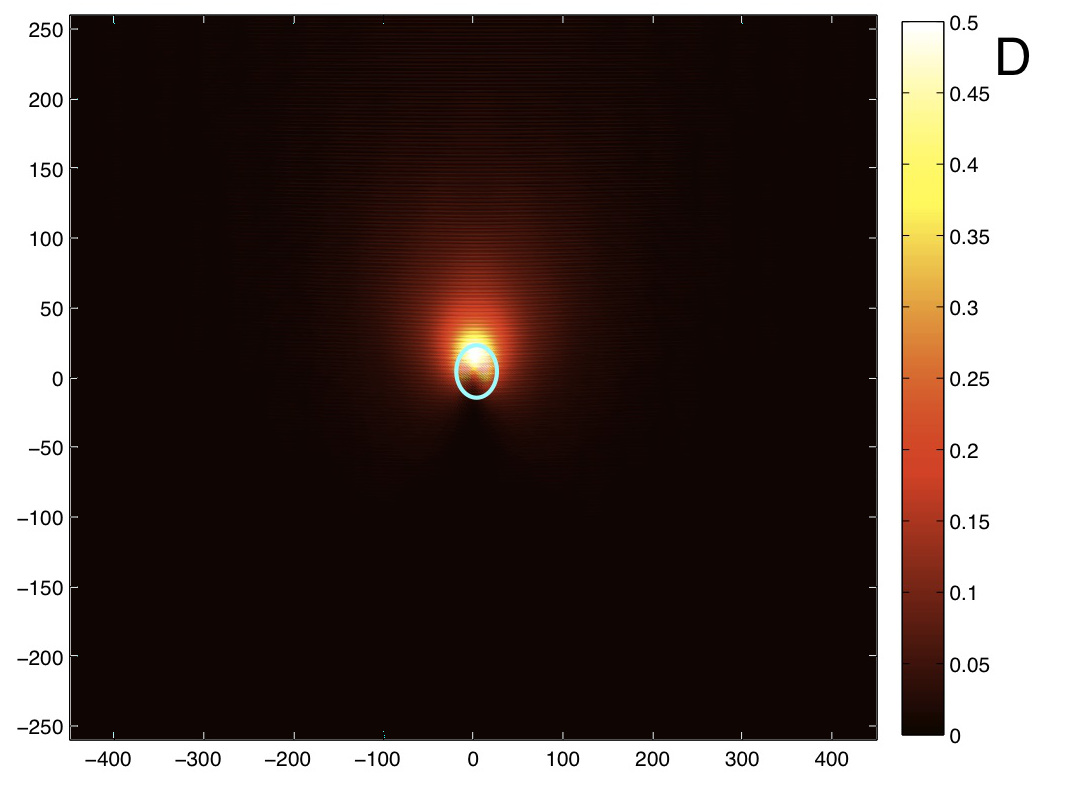}
\caption{Field intensity distribution numerically calculated from Eq.~\eqref{steadystate} for different pumping frequencies $\omega_0$. $N=601$, $\hbar\gamma/t=0.005$, $\tau=0$ and $\sigma_x/a=10$ for all panels. Panels \textbf{A} and \textbf{B} are for $\sigma_y/a=50$, while panels \textbf{C} and \textbf{D} are for $\sigma_y/a=10$. The pumping area spot is highlighted by the cyan circle. Panels \textbf{A} and \textbf{C} are for $\hbar\omega_0/t=0$, corresponding to the Dirac point, while  panels \textbf{B} and \textbf{D} are for a positive detuning $\hbar\omega_0/t=0.1$.}
\label{fig:spatialpatternzerostrain}
\end{figure} 

\subsection{Perfect honeycomb lattice}
For the unstrained case, figure~\ref{fig:spatialpatternzerostrain} shows the field amplitude $|a_{i,j}|^2$ and $|b_{i,j}|^2$ on the $A$ and $B$ sites in the steady state for two different pumping frequencies $\omega_0$ and for two different spatial extents of the pump along $y$. The first row is obtained for $\sigma_x/a=10$ and $\sigma_y/a=50$, at $\omega_0=0$ (panel~\textbf{A}) and at $\omega_0>0$ (panel~\textbf{B}). The second row is obtained for $\sigma_x/a=\sigma_y/a=10$, at the same two pumping frequencies. 

The intensity patterns shown in the figure display an interesting structure that can be qualitatively understood as follows. The pump excites waves that expand radially in all directions with the Dirac velocity for a distance of the order of $v_D/\gamma$, the so-called conical diffraction~\cite{Peleg}. The angular dependence around the pump spot is determined by the matching of the phase profile of the pump with the relative phase of the $A$ and $B$ sites at different wavevectors in the vicinity of the Dirac point. As can be seen in panels \textbf{A} and \textbf{C} for $\omega_0=0$, constructive interference reinforces the intensity in the positive-$y$ direction, while destructive interference reduces the intensity in the negative-$y$ direction. 

At finite frequency $\omega_0\neq 0$, another mechanism contributes to the determination of the pattern: as one can see in panel \textbf{B}, the intensity is now concentrated laterally in the positive and negative $x$ direction and there is almost no intensity in the positive-$y$ direction. This fact can be explained in terms of the momentum distribution of the incident field, which does not overlap with the resonant Dirac wave at a finite wavevector $k_y\simeq \omega_0/v_D\gg 1/\sigma_y$ in the positive-$y$ direction. As expected, this feature no longer occurs for a smaller $\sigma_y$ for which $k_y\simeq \omega_0/v_D<1/\sigma_y$ and a good overlap is again possible also in the positive-$y$ direction. As a result, the angular distribution is in this case again maximal in the positive-$y$ direction, see panel \textbf{D}.

\subsection{Strained honeycomb lattice}
\begin{figure*}[p!]
\hspace*{-1.5em}
\includegraphics[scale=0.45]{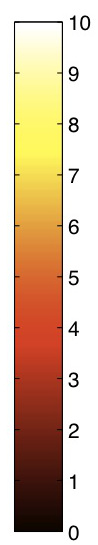}
\includegraphics[width=0.32\textwidth]{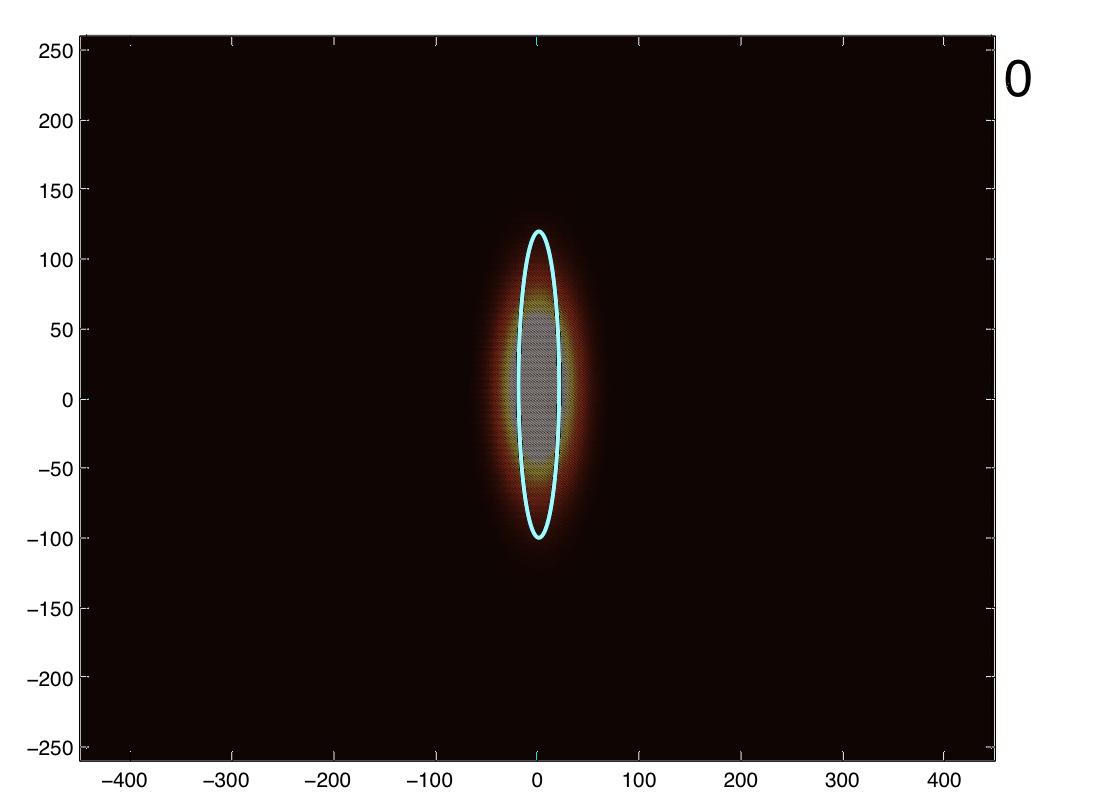}
\includegraphics[width=0.32\textwidth]{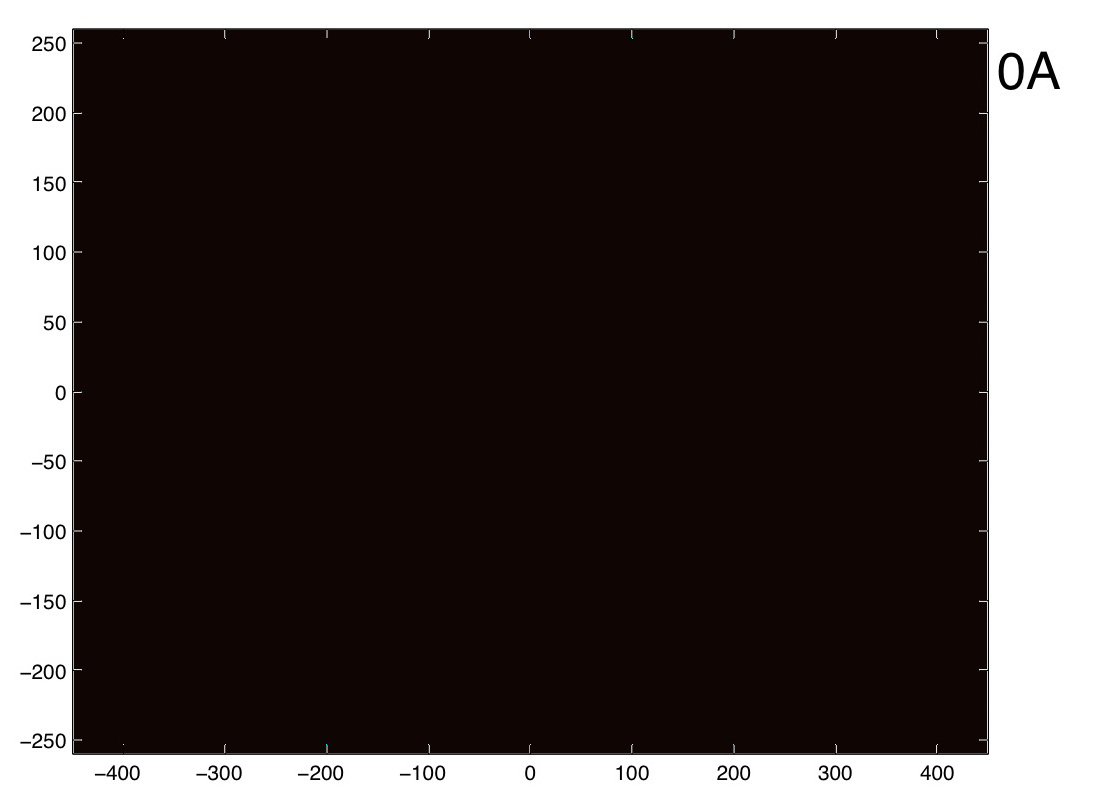}
\includegraphics[width=0.32\textwidth]{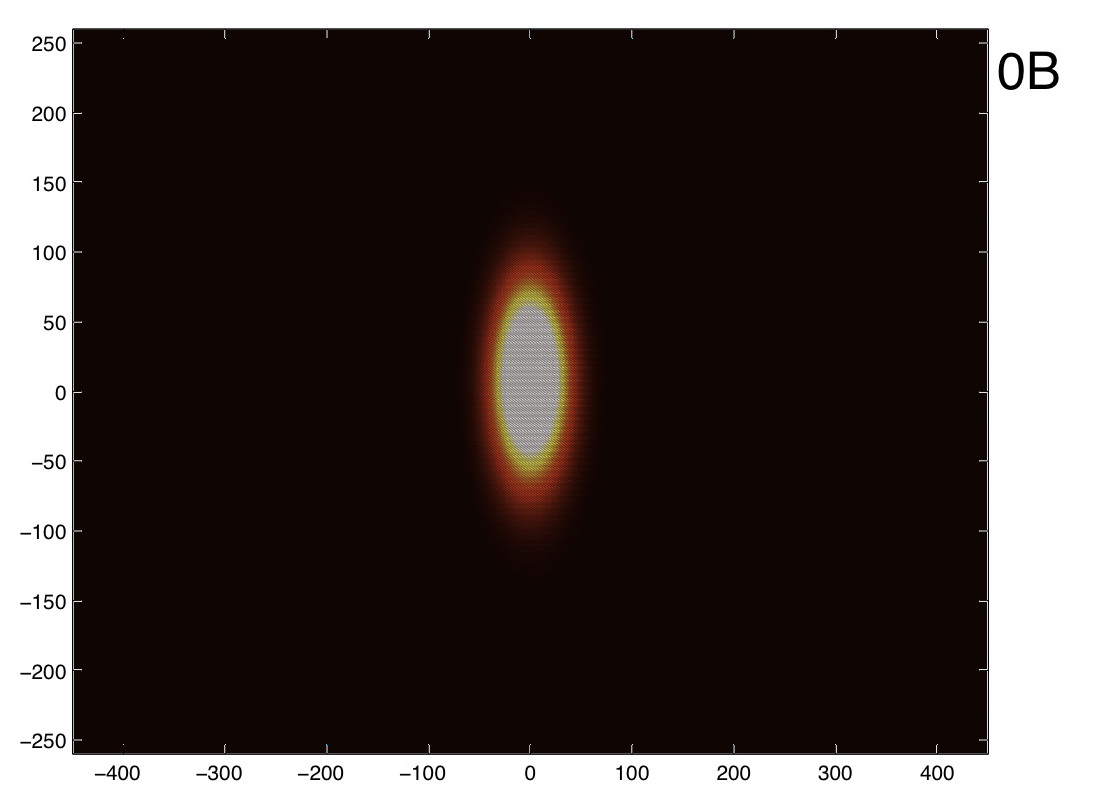}\\
\hspace*{-1.5em}
\includegraphics[scale=0.45]{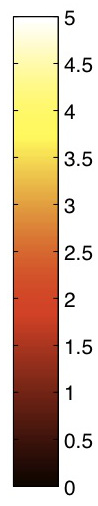}
\includegraphics[width=0.32\textwidth]{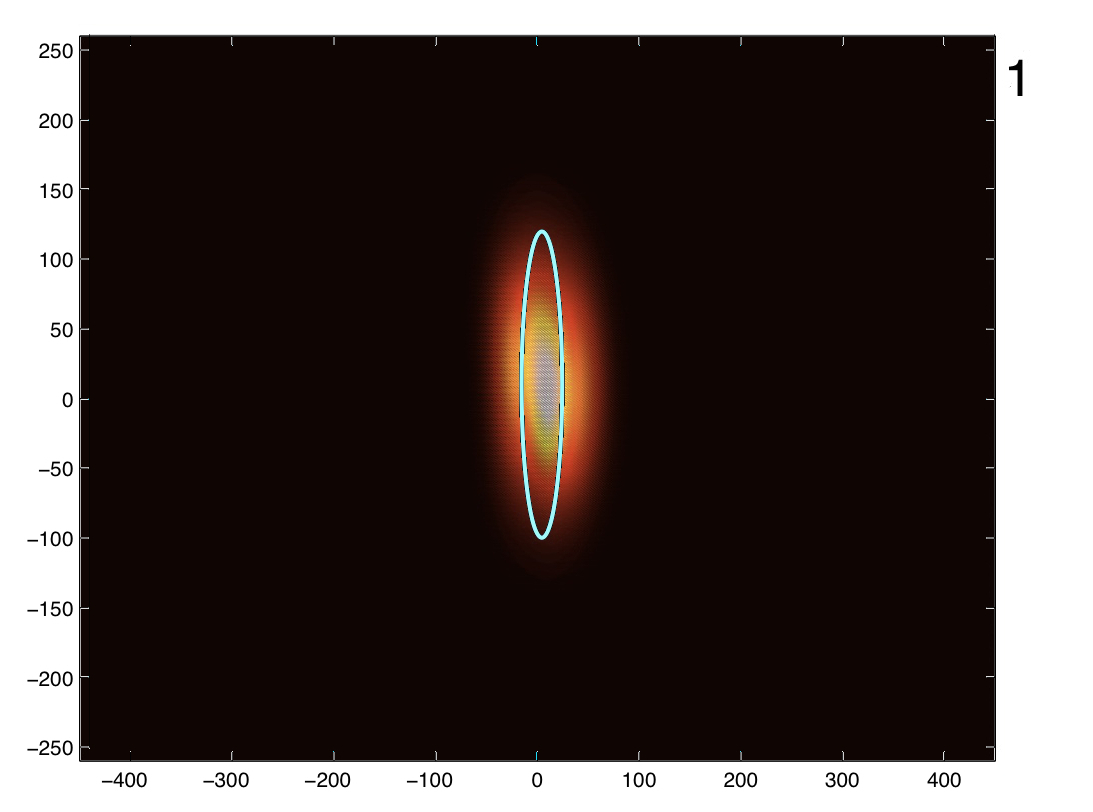}
\includegraphics[width=0.32\textwidth]{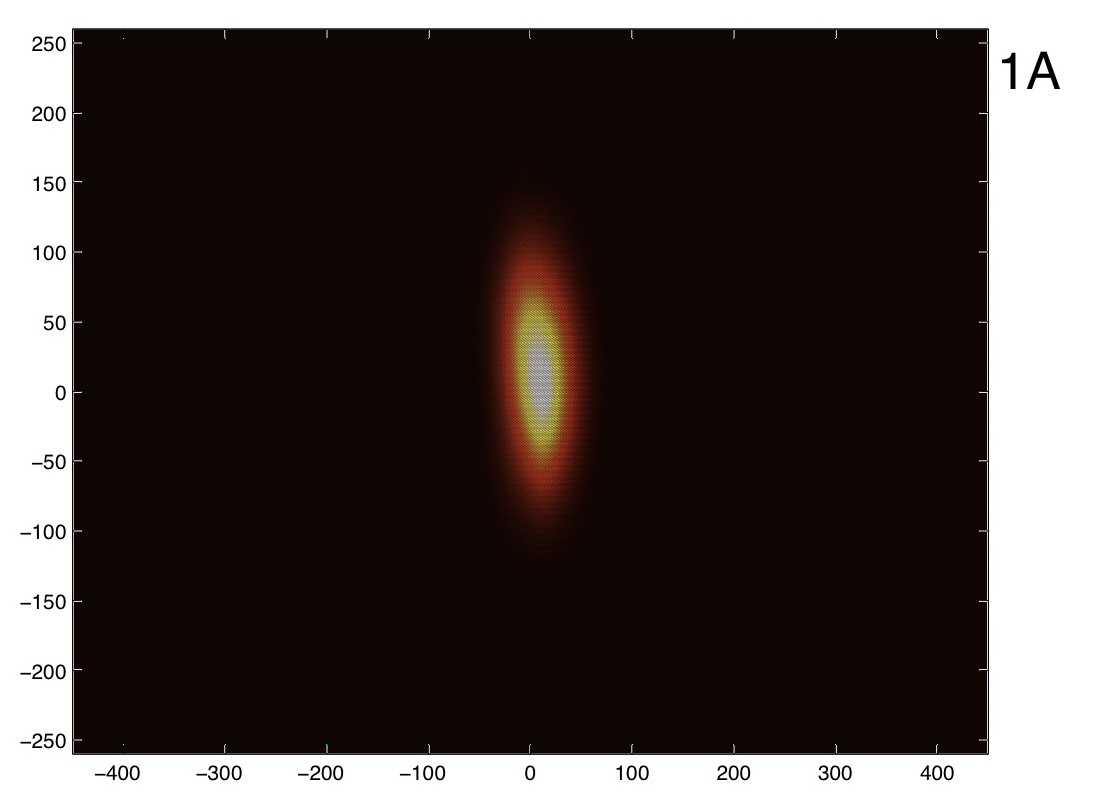}
\includegraphics[width=0.32\textwidth]{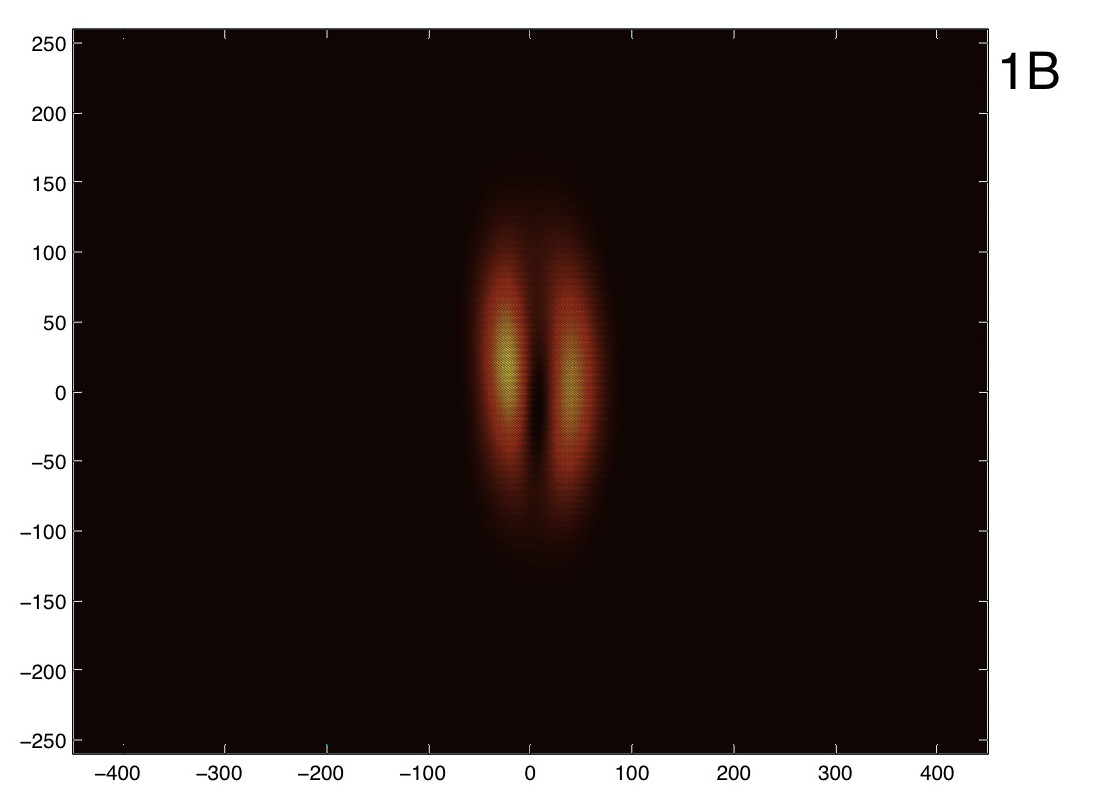}\\
\hspace*{-1.5em}
\includegraphics[scale=0.45]{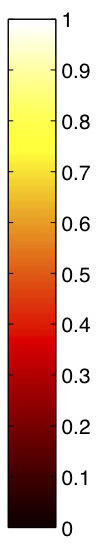}
\includegraphics[width=0.32\textwidth]{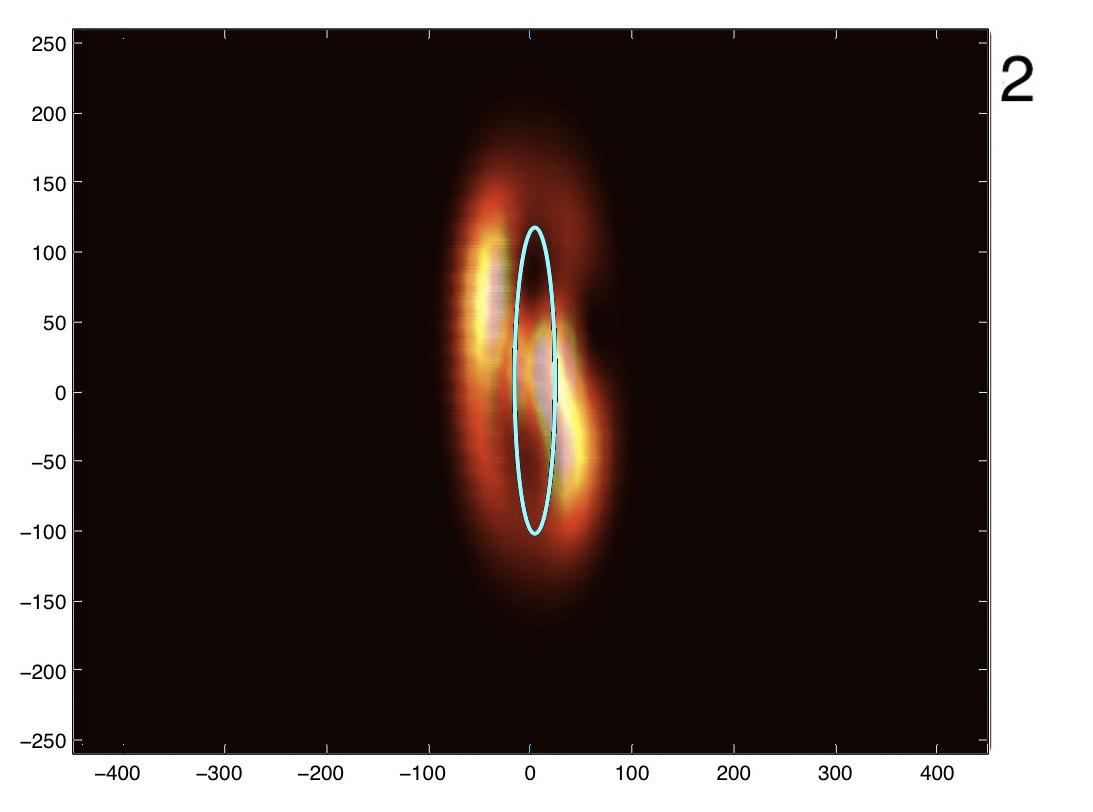}
\includegraphics[width=0.32\textwidth]{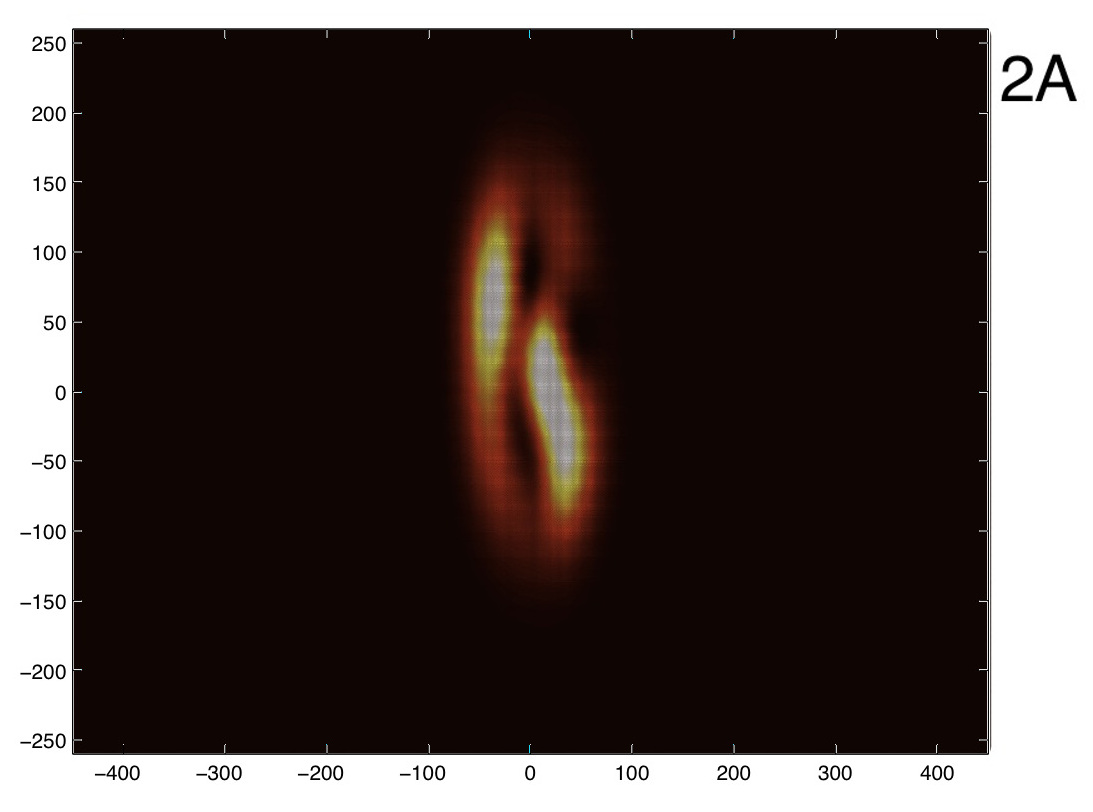}
\includegraphics[width=0.32\textwidth]{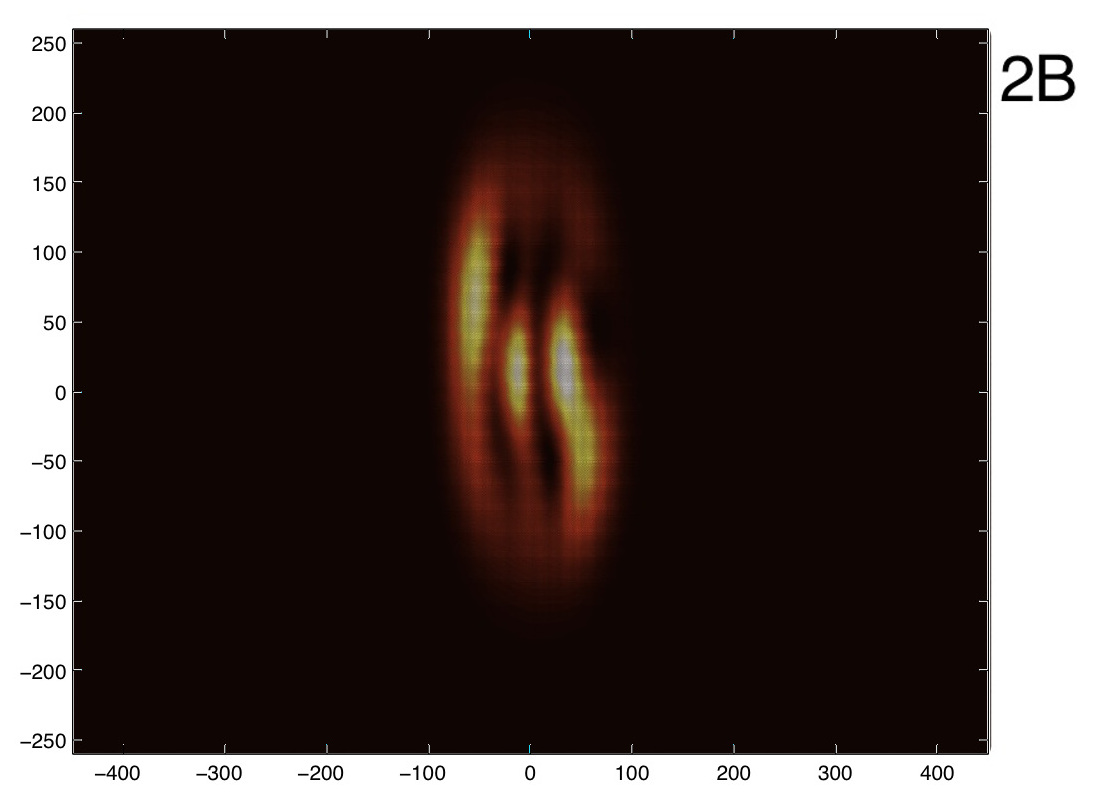}\\
\hspace*{-1.5em}
\includegraphics[scale=0.45]{colorbar_2.jpg}
\includegraphics[width=0.32\textwidth]{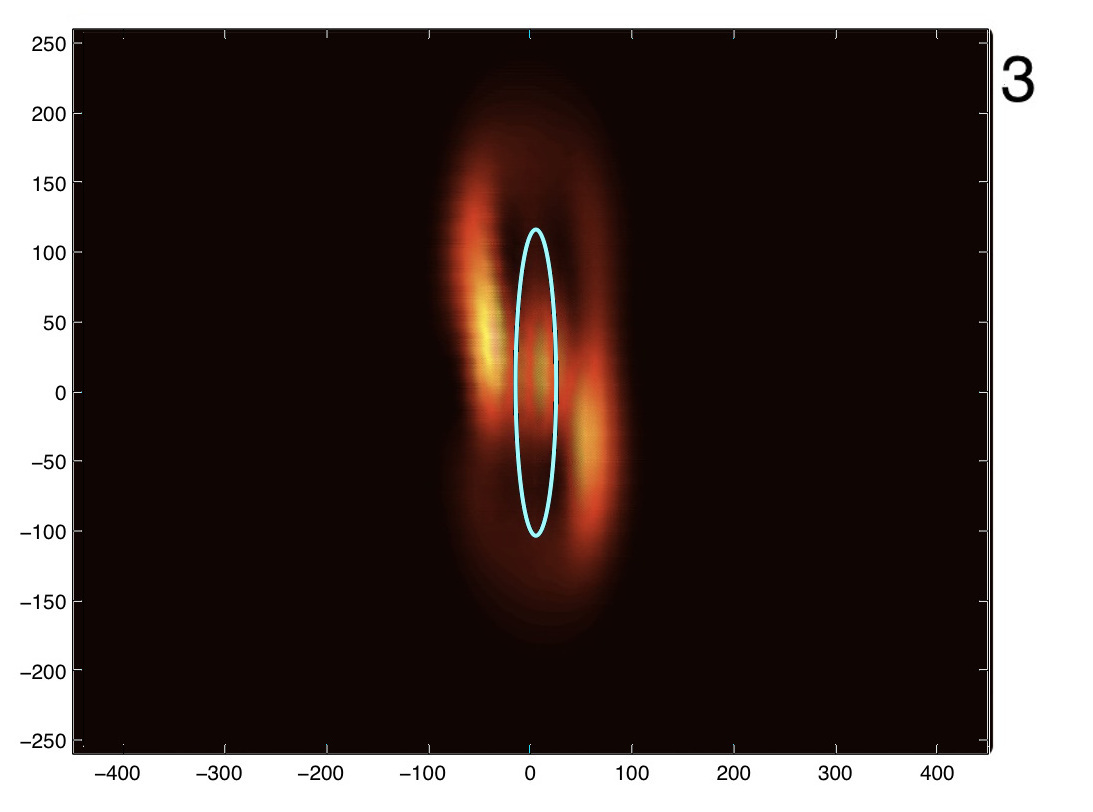}
\includegraphics[width=0.32\textwidth]{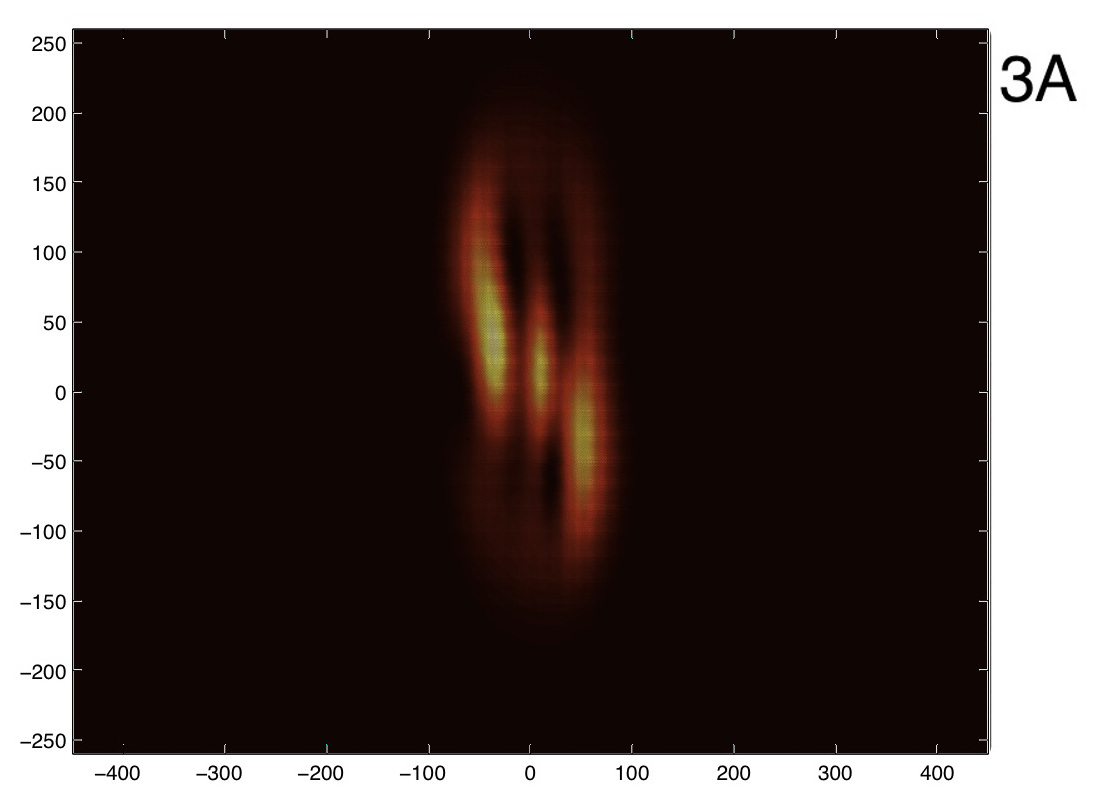}
\includegraphics[width=0.32\textwidth]{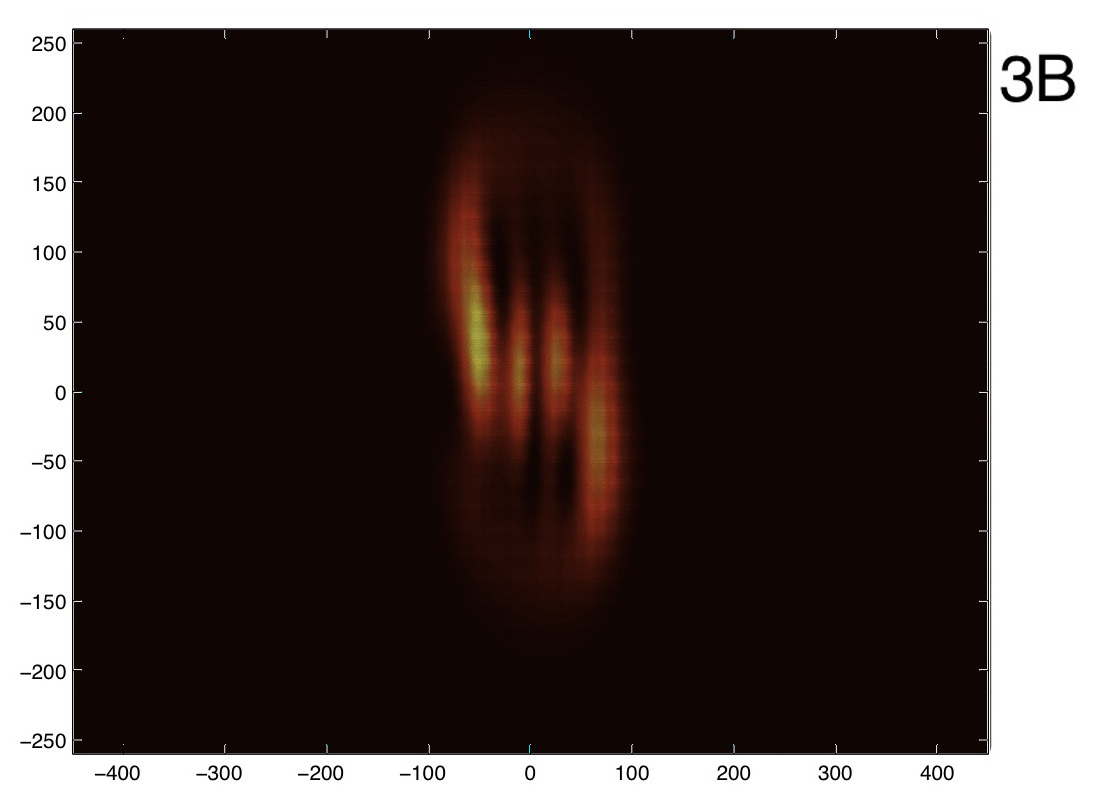}\\
\hspace*{-1.5em}
\includegraphics[scale=0.45]{colorbar_2.jpg}
\includegraphics[width=0.32\textwidth]{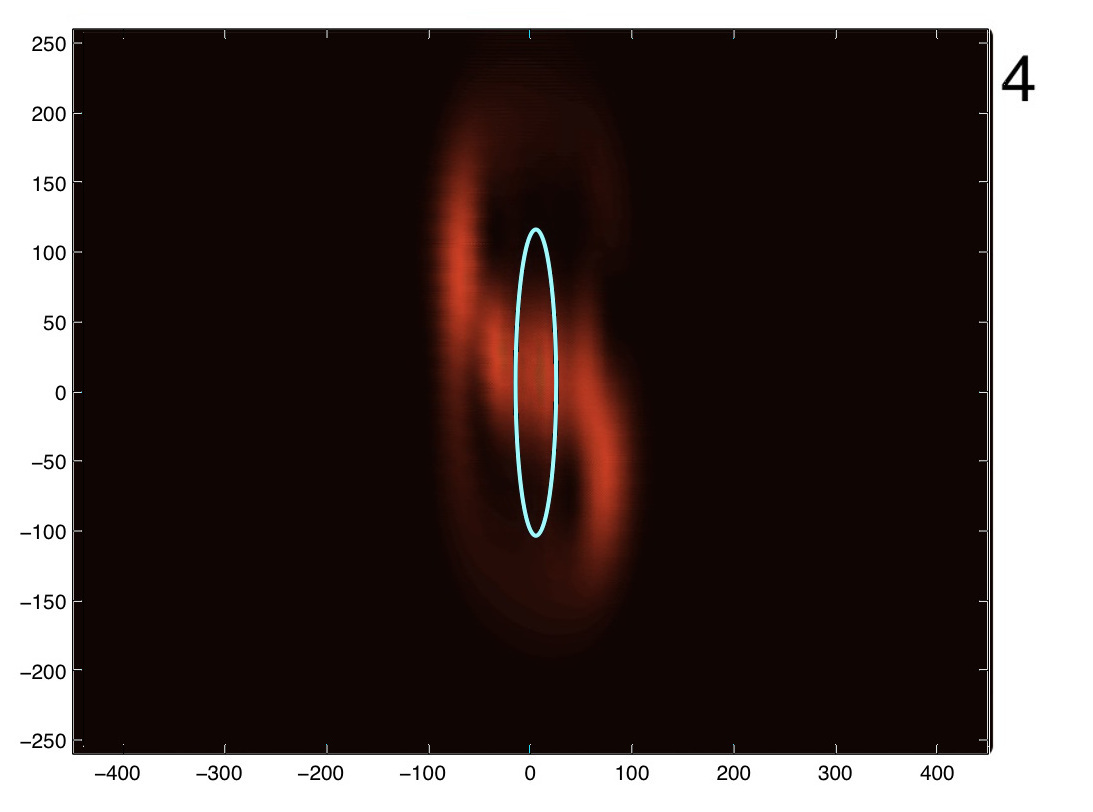}
\includegraphics[width=0.32\textwidth]{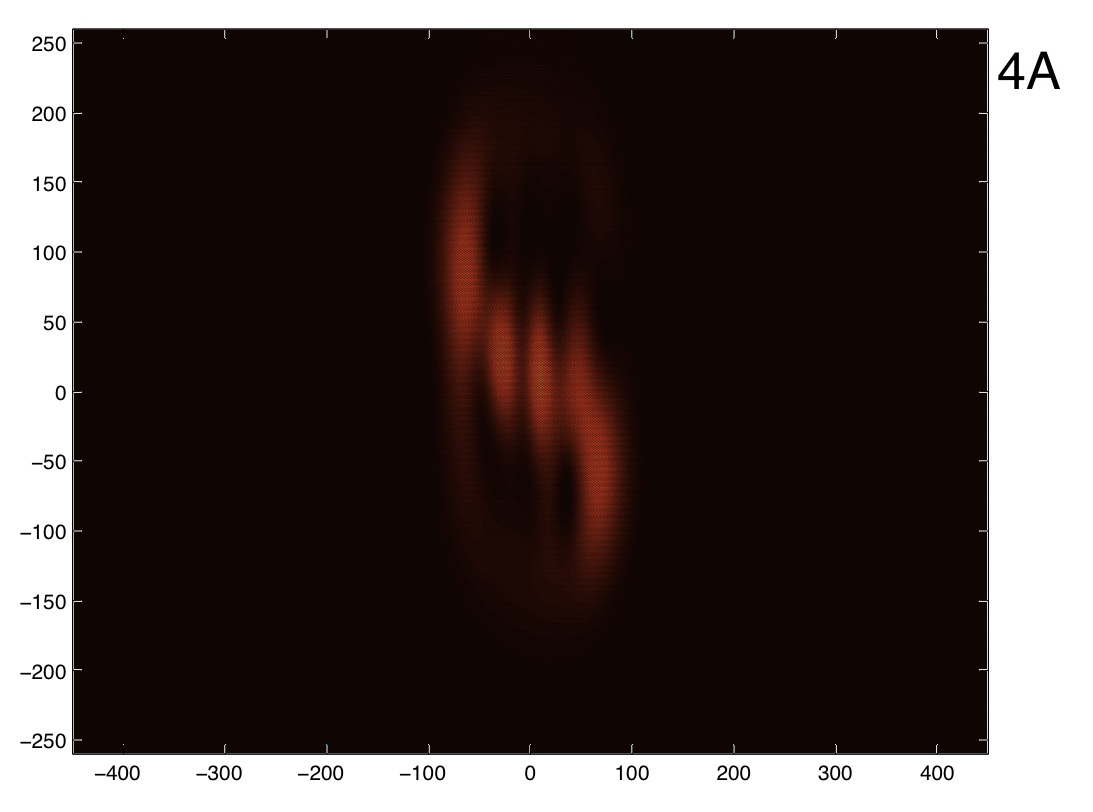}
\includegraphics[width=0.32\textwidth]{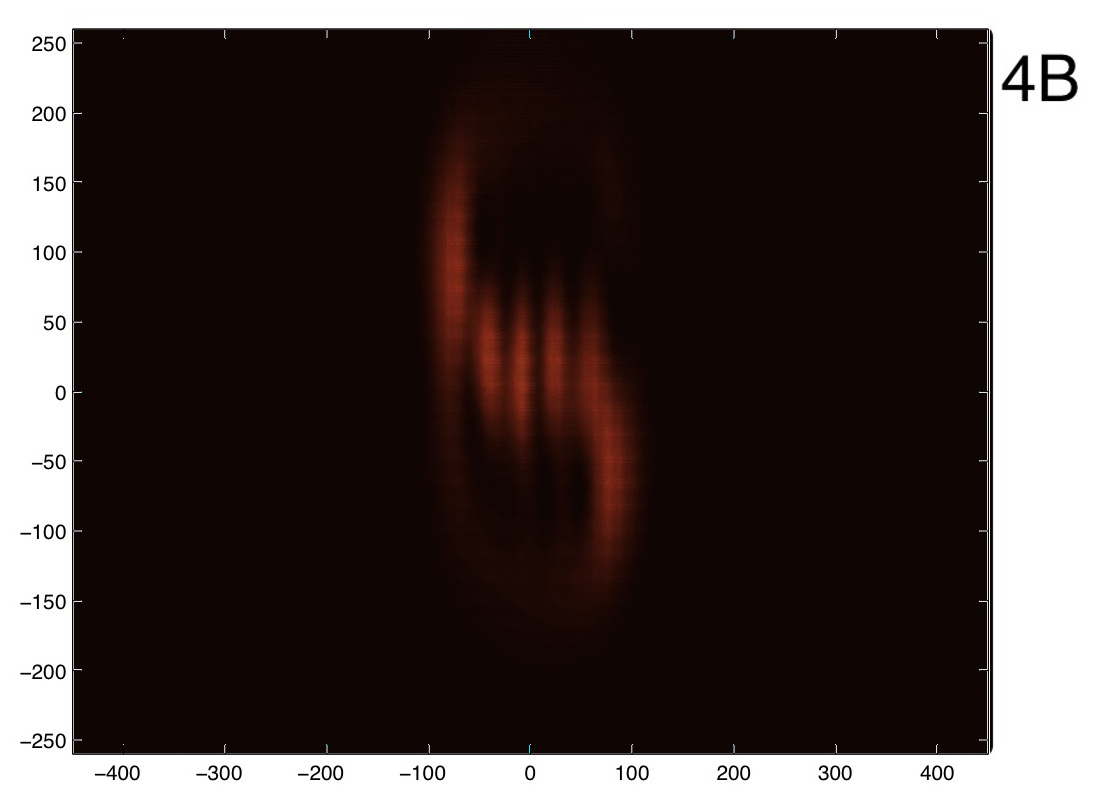}
\caption{Numerically calculated steady-state intensity distribution for different pump frequencies $\omega_0$ on resonance with the different Landau levels at the $K'$ point. We have chosen $N=601$, $\hbar\gamma/t=0.005$, $\tau=0.005$, $\sigma_x/a=10$ and $\sigma_y/a=50$ for all panels. Panels \textbf{0-4} show the total intensity for frequencies $\omega_0=0$, $\omega$, $\omega\sqrt{2}$, $\omega\sqrt{3}$, $\omega\sqrt{4}$ corresponding to the first five Landau levels. The pumped region is highlighted by the cyan circle. While the left panels show the complete intensity distribution, the central \textbf{0A-4A} and right \textbf{0B-4B} panels isolate the intensity distribution on $A$ and $B$ sites.}
\label{fig:spatialpattern}
\end{figure*} 

The spatial structure of Landau levels in a strained honeycomb lattice are illustrated in Fig.~\ref{fig:spatialpattern}: a relatively weak strain of $\tau=0.005$ is considered and the steady-states intensity patterns are shown for different pumping frequencies $\omega_0$.  Panels~\textbf{0-4} are for frequencies of the first five Landau levels $n=0\ldots 4$, according to Eq.~\eqref{ll}. Parameters for all panels were chosen to give the best representation of the Landau level wave functions with a realistic form of the pump: in particular, a relatively wide pump spot along $y$ was needed to keep the jitter of the guiding center $x_0$ smaller than the magnetic length $l_B$ so as to avoid blurring of the Landau level intensity pattern.

To better understand the mode structure, the separated field intensity distributions on respectively the $A$ and the $B$ sites are shown in the panels labelled with \textbf{A} or \textbf{B}: as expected, the qualitative structure of the intensity profile follows the shape of the eigenstates shown in Fig.~\ref{fig:n12ll}. In particular, the numbers of nodes and the width of the field amplitudes coincide with what is expected from the analytical wave functions Eq.~\eqref{statell}.

The particular spiral-like shape is due to the interference between the resonant Landau level and its neighbours, that are non-resonantly excited. By decreasing the loss rate, in the limit of $\hbar\gamma/t\rightarrow 0$, the resonant mode is more and more dominant while the non-resonant contribution is suppressed and the spiral behaviour disappears.

\section{Honeycomb lattice with next-nearest-neighbour hoppings}
\label{sec:NNN}
\begin{figure}[h]
\centering
\includegraphics[width=0.4\textwidth]{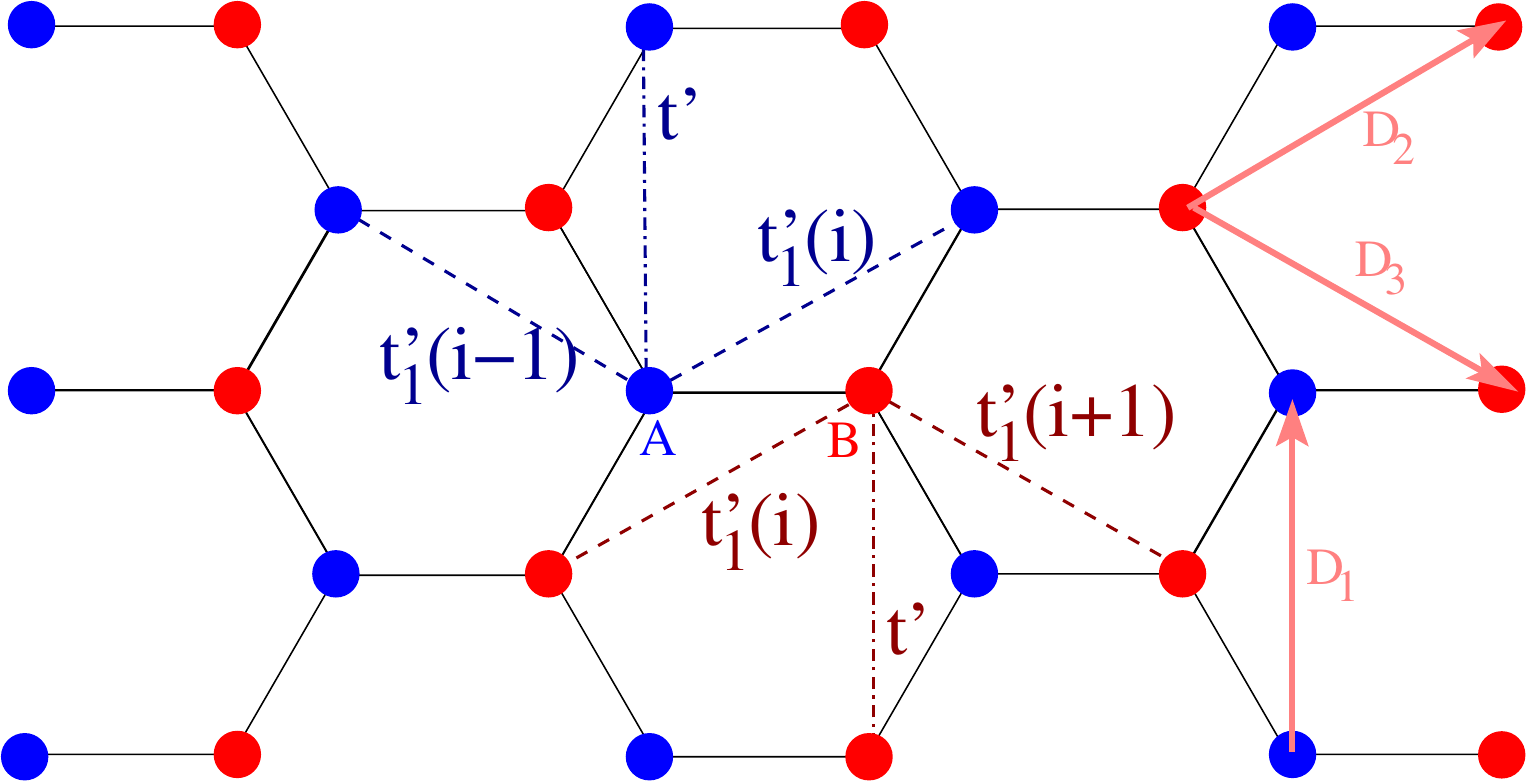}
\caption{The  honeycomb lattice with next-nearest-neighbour (NNN) hoppings. We assume that the NNN hoppings in the vertical direction, denoted by $t'$, are all equal and independent of position. The NNN hoppings along the horizontal direction $t'_1(i)$ spatially depend on the position index $i$ of the unit cell (introduced in Fig.~\ref{fig:graphene}) according to \eqref{strainNNN}.}
\label{fig:grapheneNNN}
\end{figure}
In the discussion so far, we have considered a tight-binding model of the honeycomb lattice with only nearest-neighbour hoppings, but for realistic systems, it may be important to go beyond this approximation. In real graphene, for example, electrons hop to next-nearest-neighbour (NNN) sites with an amplitude $t'$ that has been estimated as being on the order of five percent of the nearest-neighbour hopping $t$~\cite{Castroneto}. Similar estimates have also been made for the experimental realizations of an artificial honeycomb lattice, such as in~\cite{Amo1,Bellec1,Bellec2,Bellec3}. We also note that the introduction of a spatially-dependent nearest-neighbour hoppings in these experiments would result in {\it spatially-dependent} NNN hoppings. Motivated by this, we now generalize the results of the previous sections to the case of non-zero NNN hoppings.

The momentum-space Hamiltonian corresponding to Eq.~\eqref{tbh}, for a bulk system in presence of NNN hopping is: 
\begin{equation}
\ham=\begin{pmatrix}
V'_A(\vec{k})     & V^*(\vec{k}) \\
V(\vec{k}) & V'_B(\vec{k})
\end{pmatrix}=\ham_0+\ham'
\label{tbhNNN}
\end{equation}
where the anti-diagonal matrix $\ham_0$ gives the contribution only from nearest neighbour hoppings, and the diagonal matrix $\ham'$ describes the NNN hoppings, as depicted in Fig.~\ref{fig:grapheneNNN}. 
In the strained honeycomb lattice, the vertical NNN hoppings denoted by $t'$ will remain constant, while all other NNN hoppings become position dependent, as denoted by $t'_1 (i)$, where $i$ is the position index $i$ of the unit cell. Then we have:
\begin{equation}
\begin{split}
V'_A(\vec{k})&= t'\left(\e^{\ii \vec{k}\cdot \vec{D}_1}+ \e^{-\ii \vec{k}\cdot \vec{D}_1}\right) +  t'_1(i)\left(\e^{\ii \vec{k}\cdot \vec{D}_2}+ \e^{\ii \vec{k}\cdot \vec{D}_3}\right)\\& \qquad +  t'_1(i-1)\left(\e^{-\ii \vec{k}\cdot \vec{D}_2}+ \e^{-\ii \vec{k}\cdot \vec{D}_3}\right) \\
V'_B(\vec{k})&= t'\left(\e^{\ii \vec{k}\cdot \vec{D}_1}+ \e^{-\ii \vec{k}\cdot \vec{D}_1}\right) +  t'_1(i)\left(\e^{-\ii \vec{k}\cdot \vec{D}_2}+ \e^{-\ii \vec{k}\cdot \vec{D}_3}\right)\\& \qquad +  t'_1(i+1)\left(\e^{\ii \vec{k}\cdot \vec{D}_2}+ \e^{\ii \vec{k}\cdot \vec{D}_3}\right) 
\end{split}
\end{equation}
where $\vec{D}_1=(0,\sqrt{3}a)$, $\vec{D}_2=(3a/2,\sqrt{3}a/2)$, $\vec{D}_3=(3a/2,-\sqrt{3}a/2)$ are the NNN lattice vectors. We now also assume that the $t'_1(i)$ hopping has the same spatial dependence as the nearest-neighbour hoppings in Eq.~\eqref{strain}, \textit{i.e.}:
\begin{equation}
t'_1(i)=t' \left(1+\frac{x_i}{3a} \tau \right)
\label{strainNNN}
\end{equation}
Expanding to first order in small $|\vec{q}|\ll 1/a$ around the Dirac point, as done previously in Sec.~\ref{subsec:strain}, we get:
\begin{equation}
\begin{split}
V'_A(\vec{q})&\approx -3 t' -t'\tau\left(\frac{2}{3a} \hat{x} -1\right)\\
V'_B(\vec{q})&\approx -3 t' -t'\tau\left(\frac{2}{3a} \hat{x} +1\right)
\label{correctionNNN}
\end{split}
\end{equation}
where we have assumed that the spatial dependence is weak such that $t'_1(i)-t'_1(i+1)\ll t'$; this condition is guaranteed when $\tau \ll 1$ as we have assumed throughout this paper.
Taking $t'\ll t$, we estimate the energy correction due to the NNN hoppings using first order perturbation theory as
\begin{equation}
\Delta \mathcal{E}= \begin{pmatrix} \phi_A & \phi_B \end{pmatrix}  \ham' \begin{pmatrix} \phi_A \\ \phi_B \end{pmatrix} = - 3t' - \frac{2t'\tau}{3a} x_0
\label{correction2NNN}
\end{equation}
where we have used that $\hat{x}=x_0 + \sqrt{\frac{\hbar}{2m\omega}}(\hat{a}+\hat{a}^\dagger)$ and that $\phi_A$ and $\phi_B$ are eigenfunctions of the number operator $\hat{N} = \hat{a}^\dagger \hat{a}$.

Including the NNN hopping correction, the Landau level energy spectrum is now $\mathcal{E}_n^\text{NNN}=\Delta \mathcal{E}+\mathcal{E}_n$:
\begin{equation}
\mathcal{E}_n^\text{NNN}=-3t' - 3 \xi t' q_y a \pm t \sqrt{\tau |n|}\sqrt{1-\xi q_y a}
\label{llcorrectionNNN}
\end{equation}
where we have used Eq.~\eqref{shift} to evaluate the oscillation center $x_0$ in Eq.~\eqref{correction2NNN}. We notice that exactly at the Dirac point, the effect of the NNN hoppings is only to contribute a global energy shift.
\begin{figure}[]
\hspace*{-1em}
\includegraphics[width=0.51\textwidth]{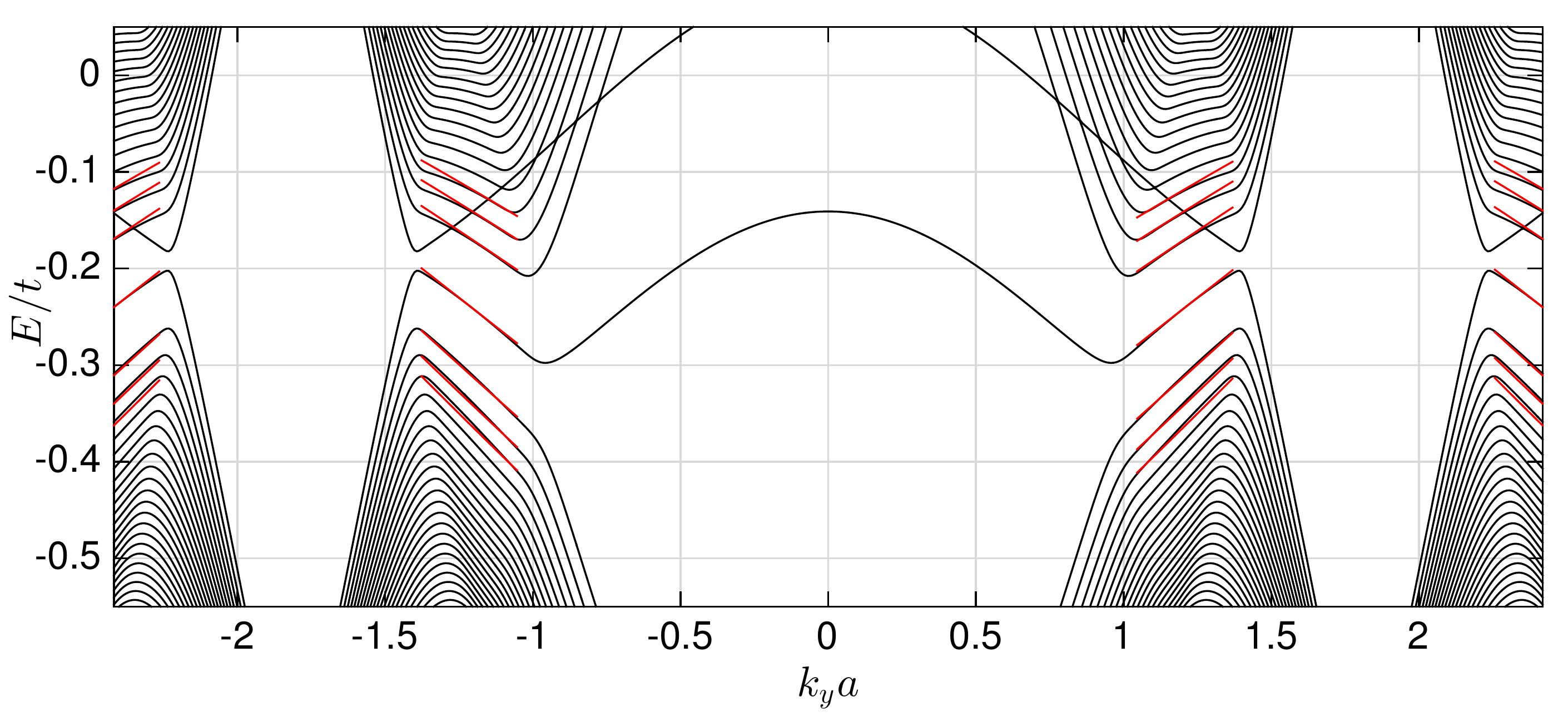}
\caption{The structure of levels around zero energy as a function of $k_y$, in units of the bare hopping $t$. This is numerically calculated from exact diagonalization of the tight-binding Hamiltonian for a ribbon of $N_x=601$ unit cells along $x$ with periodic boundary conditions along the $y$-direction, including next-nearest-neighbour hoppings with $t'=0.08t$ and strain $\tau=0.005$. Red lines indicate the analytical prediction for the lowest Landau levels according to Eq.~\eqref{llcorrectionNNN}.}
\label{fig:llNNN}
\end{figure}  

In Fig.~\ref{fig:llNNN} we plot the structure of low-energy levels as numerically obtained from exact diagonalization of the tight-binding Hamiltonian for the same parameters as in Fig.~\ref{fig:ll}, supplemented by a non zero NNN hopping $t'=0.08 t$. Numerical data are compared to the analytical prediction given by  Eq.~\eqref{llcorrectionNNN}: good agreement is clearly visible in the regions around the $K$, $K'$ points. We have also observed that all qualitative features of the wavefunction remain unchanged by the addition of the NNN hoppings. 

\section{Experimental remarks}
\label{sec:experim}
From the experimental point of view, the requirements to clearly observe the patterns discussed in the previous section are a relatively small value of the loss rate and a relatively large size of the lattice. On the one hand, stronger loss rates may hinder a clear identification of the Landau level peaks in the spectra of Fig.~\ref{fig:spectra} and are responsible for strong mixing of neighbouring eigenstates in the spatial pattern of Fig.~\ref{fig:spatialpattern}. 

On the other hand, in smaller lattices spurious reflections at the lattice edges may occur as well as significant distortions of the mode wavefunctions. While propagation towards the lattice edges can be a problem for propagating Dirac waves in perfect honeycomb lattices, this issue is much less severe in the presence of strain as the Landau levels are spatially localized. As a result, finite size effects are negligible as soon as the size of the lattice is larger the magnetic length $l_B$.

In particular, we note that two consecutive Landau levels can be resolved if the separation between them is larger than the linewidth, namely $\hbar\gamma < \mathcal{E}_n-\mathcal{E}_{n-1}$. This means that the $n$-th gap is resolved if: 
\begin{equation}
\frac{\hbar\gamma}{t}< (\sqrt{n}-\sqrt{n-1})\frac{\sqrt{\tau}}{2}.
\label{cond1}
\end{equation}

In order for the Landau wave function not to be distorted by the edges of the lattices, we need the magnetic length to be much smaller than the size of the system. The total length of the lattice in Fig.~\ref{fig:graphene} is $L_x=(3N_x-1)a/2$. The condition $l_B \ll L_x/2$, then, implies that: 
\begin{equation}
\tau > \left(\frac{6\sqrt{2}}{3N_x-1}\right)^2.
\label{cond3}
\end{equation}
In order for the key features of Landau level wavefunction not to be blurred, such as the number of nodes or the width of the wave function, one also needs that the position $x_0$ of the guiding center jitters by less than a magnetic length under the uncertainty of $q_y\simeq 1/L_y$: remarkably, this imposes a condition $l_B \ll L_y$ analogous to~\eqref{cond3}.
Finally, we need to keep in mind the condition already given in Eq.~\eqref{cond0}, which set an upper bound on the strain $\tau$ in order to have a physical $t_1\geq 0$ hopping at all points of the lattice:
\begin{equation}
\tau < \frac{12}{3N_x-1}.
\label{cond4}
\end{equation}

\begin{figure}
\includegraphics[width=0.35\textwidth]{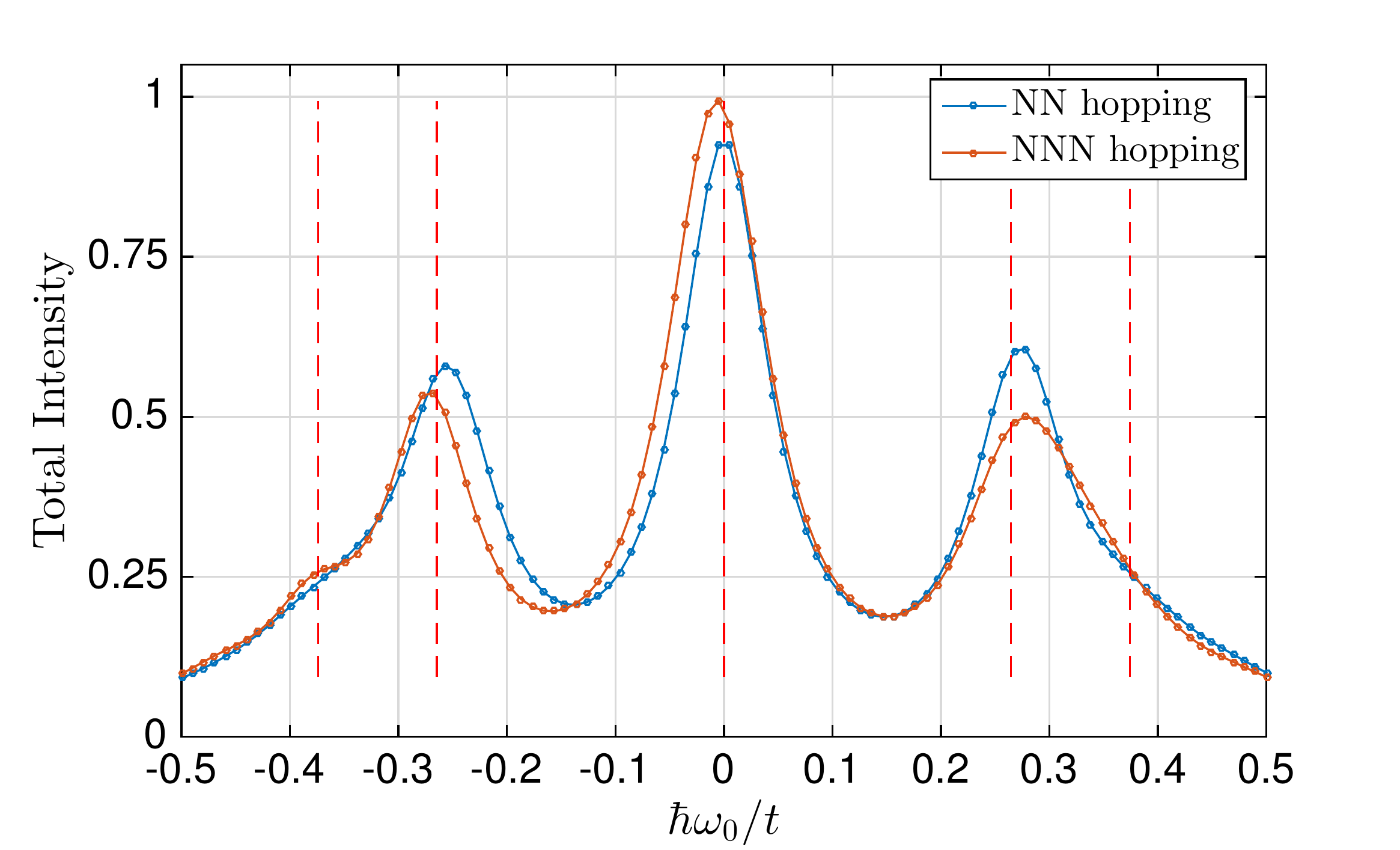}\\
\includegraphics[width=0.235\textwidth]{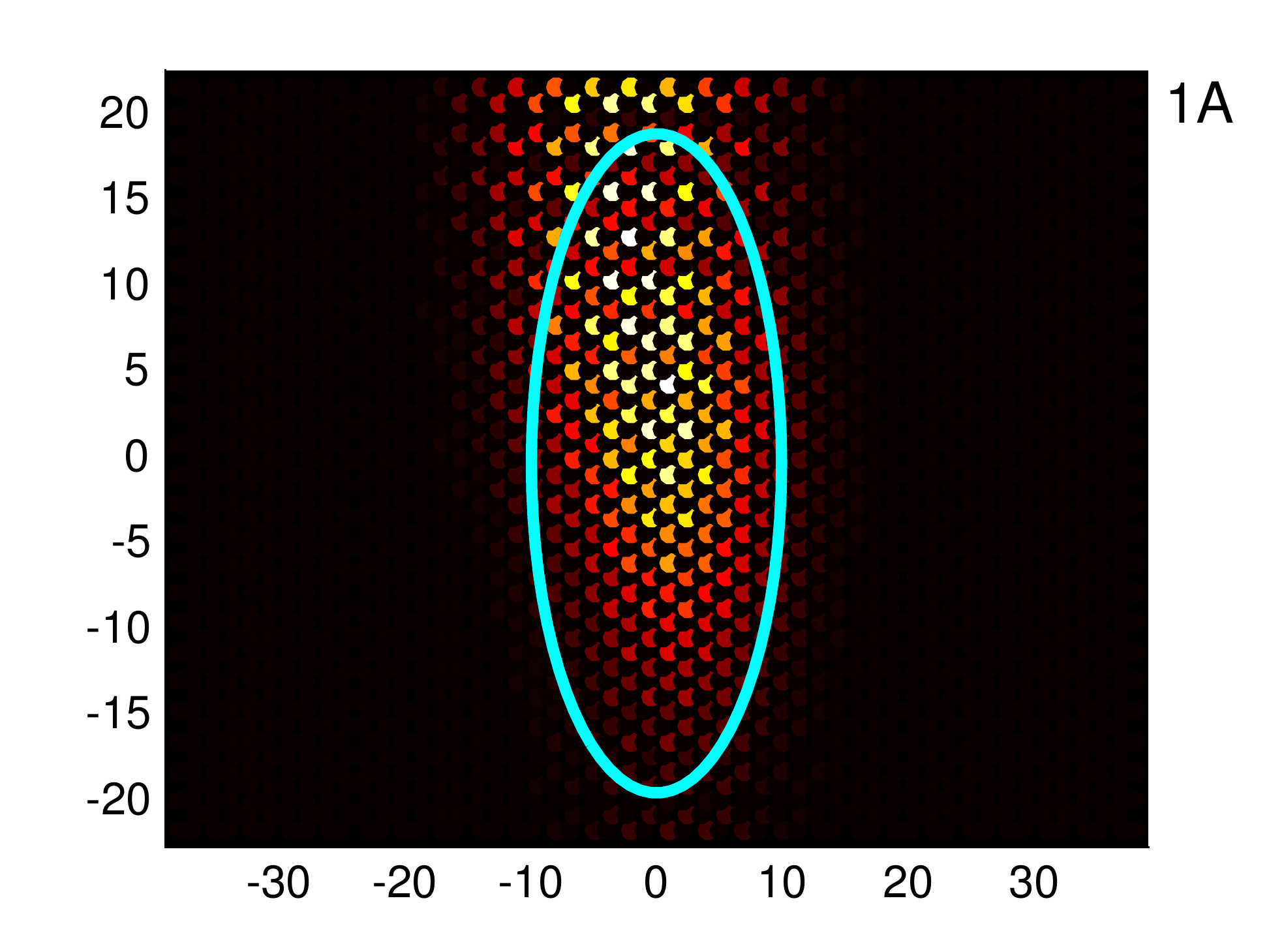}
\includegraphics[width=0.235\textwidth]{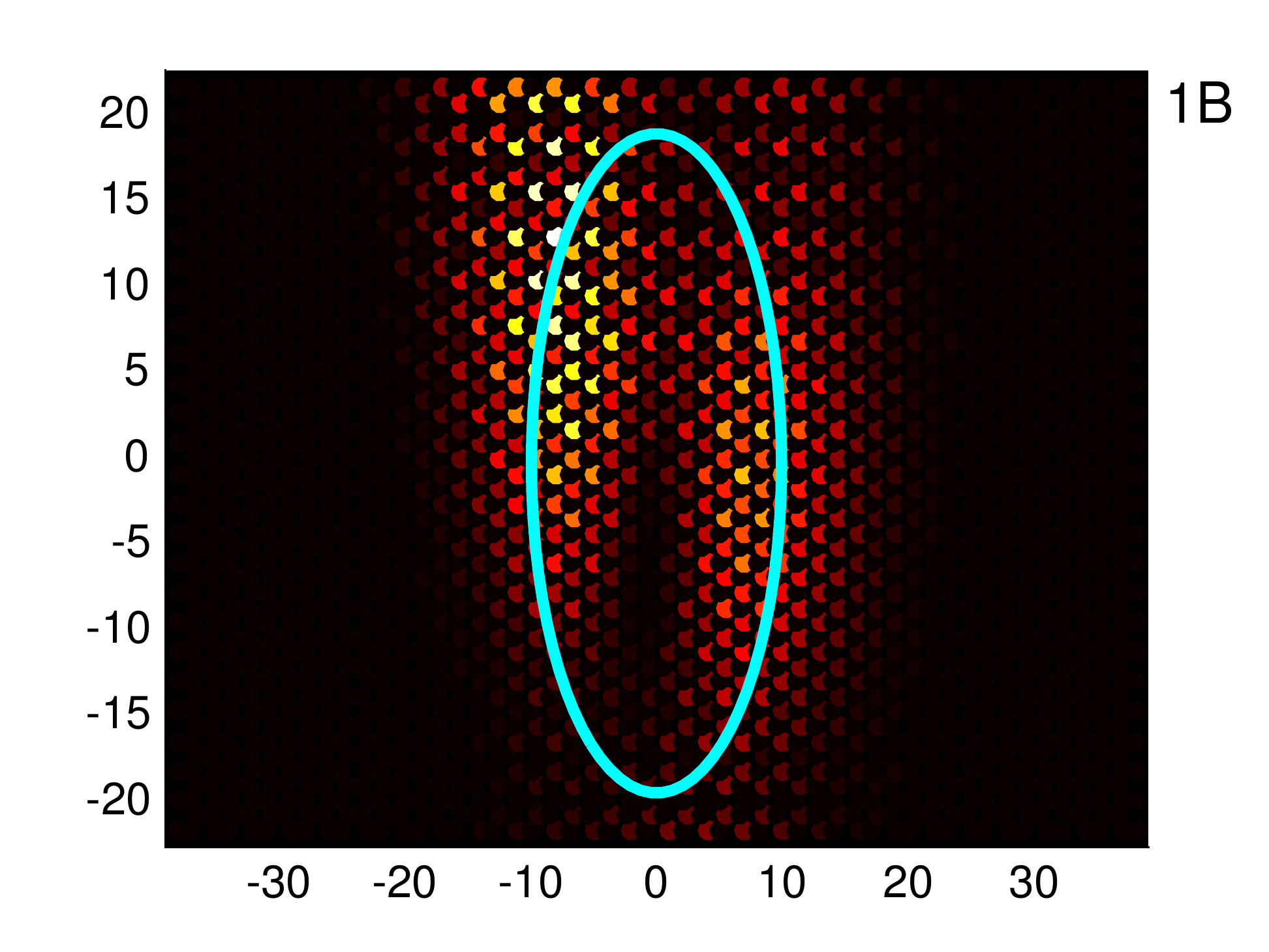}
\caption{Numerical calculations using realistic parameters of state-of-the-art experimental setups, namely $N_x=N_y=51$, $\hbar\gamma/t=0.05$, $\tau=0.07$, $\sigma_x/a=5$ and $\sigma_y/a=10$. Upper panel: spectra of the total field intensity as a function of the pumping frequency $\omega_0$. The blue curve is the spectra obtained with only nearest-neighbour hopping, the orange curve includes contributions also from the position dependent next-nearest-neighbour (NNN) hopping, with $t'=0.08t$. The NNN spectra has been globally shifted by the factor $-3t'/t$  from Eq.~\eqref{correction2NNN} in order to directly compare it with the spectra obtained with only nearest-neighbour hoppings. Solid lines are a guide to the eye and red dashed lines are the analytical energies of the first three Landau levels as predicted by Eq.~\eqref{ll}. Lower panels: steady-state intensity distributions on the $A$ (panel 1\textbf{A}) and $B$ (panel 1\textbf{B}) sites for a pump frequency $\omega_0$ tuned on resonance with the $n=1$ Landau level at the $K'$ point, in the presence of NNN hopping processes with a realistic amplitude $t'=0.08t$.}
\label{fig:small}
\end{figure}

For the sake of concreteness, we can discuss these criteria having in mind a realistic experiment using photonic~\cite{Amo1} or microwave~\cite{Bellec1,Bellec2,Bellec3} technology: state-of-the-art samples are in both cases restricted to relatively small lattices, with a few tens of sites in each direction. From Eq.~\eqref{cond4}, this imposes an upper bound to the strain $\tau < 0.08$. 

For a realistic loss rate $\hbar \gamma/t\approx 0.05$ of current experiments, a value $\tau \approx 0.07$ of the strain should however allow us to resolve the first two gaps between Landau levels. This is illustrated in the upper panel of Fig.~\ref{fig:small}, where we show the total intensity spectra as a function of the pump frequency: peaks corresponding to the lowest Landau levels are clearly visible with an excellent agreement with the analytical prediction of Eq.~\eqref{ll}. We also show that the spectra are only slightly affected when a position-dependent NNN hopping \eqref{strainNNN} is included with a realistic amplitude $t'=0.08t$.

In the lower panels of Fig.~\ref{fig:small} we show the intensity distribution for a pump on resonance with the peak corresponding to the $n=1$ Landau level. Independently of the presence of a NNN hopping, the peculiar nodal profile of the mode is clearly visible as a central black stripe in the $B$ sites intensity pattern shown in panel~\textbf{B}. The horizontal dark fringes that are visible in the upper and lower part of the image are, instead, a spurious effect due to reflections on the edges of the lattice.

\section{Conclusions}
\label{sec:conclu}
In this work we have proposed an experimentally-viable spectroscopic method to study Landau levels of a Bose field in a coherently-driven dissipative honeycomb lattice in the presence of an artificial pseudo-magnetic field. Focussing on an inhomogeneous unidirectional strain configuration, we have shown how the Landau levels are clearly visible as peaks in the absorption/transmission spectra of the device. When the pump frequency is tuned in the vicinity of a peak, the intensity distribution in the lattice provides direct information on the microscopic structure of the Landau level wavefunction.

Experiments along these lines appear feasible with state-of-the-art technology using photonic cavity array devices in either the infrared or microwave domains. In addition to illustrating the physics of Landau levels for relativistic massless Dirac particles, such experiments would open the way to studying rich wave-propagation physics in distorted honeycomb lattices, including Klein tunneling~\cite{Tomoki_new} and, on more speculative grounds, relativistic~\cite{Iorio} phenomena.

\section*{Acknowledgements}
We are grateful to A. Amo, P. Andreakou, M. Mili\'{c}evi\'{c} and M. Bellec, for continuous exchanges. This work was supported by the ERC through the QGBE grant, by the EU-FET Proactive grant AQuS, Project No. 640800, and by the Autonomous Province of Trento, partially through the project ``On silicon chip quantum optics for quantum computing and secure communications'' (``SiQuro'').

\end{document}